\documentstyle[aps,eqsecnum]{revtex}
\begin{document}
%\baselineskip=16pt 
%\draft
\title{\hspace{10cm} Preprint WSU-NT-14-1996 \\
\vspace{1cm}
 QCD evolution with longitudinal fields and heavy quarks} 
\author{ A.  Makhlin and  E. Surdutovich}
\address{Institute for Nuclear Theory, University of Washington,
Seattle, WA 98195 \\ and \\
Department of Physics and Astronomy, Wayne State University, 
Detroit, MI 48202}
\date{October 31, 1996}
\maketitle

\begin{abstract}

QCD evolution equations  that naturally include longitudinal (non-propagating)
fields and heavy quarks are derived. We start with the integral equations of
quantum field kinetics and obtain the master equations, similar to DGLAP
evolution equations after several consecutive approximations. We demonstrate
that in their primary form, the evolution equations include a new element,
feed-back via longitudinal fields, leading to a low-$x$
enhancement in the {\em e-p} DIS cross section.  We show that the structure
function $F_L$ is very sensitive to the dynamics of the longitudinal fields 
and that the heavy quarks in evolution equations make this effect even more
pronounced.

\end{abstract}
%\pacs{12.38.Mh, 12.38.Bx, 25.75.+r}
% body of paper here

\section{Introduction}\label{sec:SN1} 

In this paper we address  the dynamics of  longitudinal fields and  heavy
quark fields in high-energy processes, keeping in mind practical importance
of these questions for the theory of ultra-relativistic heavy ion
collisions. One of the main difficulties  of this (not yet existing) theory is
the  impossibility of describing the initial state (two nuclei), the transient
process (which, hopefully, includes the QGP stage) and the hadrons of
the final state using the standard language of scattering theory. In fact,
even in the much simpler problem of the $pp$-collisions, one has to make 
certain
model assumptions (like, {\em i.g.},  the parton  model) in order to put the
problem in the form required by the scattering theory. Unfortunately, for 
nuclear collisions, the situation is much worse, since one cannot even think
about consistent decomposition of the whole  process into the hard and the
soft parts; the factorization scheme is inapplicable.  The most unfortunate
circumstance is that the structure functions of the protons, the partonic
substitutes for the unknown wave functions, cannot be used any more.  

Two possible ways to cure the problem were suggested recently. McLerran
and Venugopolan have started with a classical model of a large nucleus in the
infinite  momentum  frame (IMF) \cite{Raju}. This model has been gradually 
improved  by  Jalilian-Marian, Kovner, McLerran, Venugopolan and Weigert 
by accounting for the small quantum fluctuations in the strong "external"
classical field \cite{Raju95,Larry}. A remarkable yield of this
ongoing study is an understanding that the  so-called low-$x$ enhancement may
be an effect of  the classical field itself \cite{Larry1}.

The idea of a geometrical and dynamical similarity between all  deeply
inelastic high energy processes, including the inelastic high energy
$ep$-scattering, is the guide for our study \cite{QFK,QGD}. The method of 
Quantum Field Kinetics (QFK) \cite{QFK} allows one to show that  the dynamics
of all inclusive processes is sequential on the real time scale and ends
at the moment of  measurement. Consequently, one can avoid 
factorization in the derivation  of the evolution equations  \cite{QGD}. The
best prospect of this approach is connected with the hope of extracting from the
DIS data more rich information ({\em e.g.}, information about the dynamics
of the static fields,) than is usually done with the aid of the DGLAP
evolution equations of the parton model  \cite{AP,Lip,GD}. The causal
character of the QCD evolution which is clearly visible due to the QFK
technique does imply such an opportunity.  As a result, one may treat the
hadronic collision not as the scattering of partons, but  as the
interaction of the fields from the colliding hadrons or nuclei. In this paper,
we continue the study of Ref.\cite{QGD} and show that the presence of the
static (not propagating) fields in the evolution equations leads to a steep
power-like behavior of the inelastic $ep$-cross section at low $x$.

The QFK approach to deep inelastic processes allows us to address the
question of the accuracy of the evolution equations as well. 
The complexity of this
problem in the calculations based on the operators product expansion (OPE) is
well understood. Ranking the operator functions by their twists and the
coefficient functions by the order of their perturbative expansion, the OPE
employs two essentially different expansions, and their interplay is difficult
to keep under the control (see, {\em i.g.}, Ref. \cite{Martin}). Our approach
is different; we descend from the Schwinger-Dyson equations all the way to the
DGLAP equations and study what of physical importance is lost on this way.  We
show that the dynamics of the classical (longitudinal) fields is an unavoidable
partner of the propagating (transverse) fields in the QCD evolution.   As a
result, the equations for the transverse fields acquire a kind of feed-back via
the inherently  inseparable longitudinal fields. By examination, 
this new element
may alone lead to the low-$x$ enhancement of the inclusive cross-section. One 
 does not even need to have a classical source like the valence quarks of the
McLerran-Venugopalan model. In order to obtain the master equations, like
DGLAP, one has to exclude the longitudinal fields from the evolution equations
by the brute force. We argue that this step is equivalent to the introduction
of the parton picture. It may be consistent only if the so-called factorization
scale, ~$\mu_{0}^{2}$,~ is a measured  parameter. 

In their full form, the evolution equations are nonlocal in the transverse
direction. In conjunction with the collinear problems of  null-plane
dynamics, this leads to severe infrared singularities, which are easily
regulated physically, but at the price of abandoning the null-plane
dynamics as a technical tool.  In order to obtain the master equations,
any evidence of the non-locality has to be eliminated, once again,
by a brute force.  Thus, we solve (or at least address) the problem of the
accuracy of the DGLAP equations by the extension of the system of evolution
equations beyond their standard mathematical form and by a revision of
their physical content. 

The way we derive and study the evolution equations allows us to incorporate
the massive quark fields into the  QCD evolution. This is  a very real
problem  since  a vast set of the current and forthcoming experiments requires
a theoretical understanding of the role of  heavy quarks in high energy
processes. At HERA energies, the range of the accessible momentum transfer
between the electron and the proton is very wide and the $c$- and $b$-mesons
are quite frequently met among the secondaries, affecting the measured cross
section of deep inelastic scattering (DIS). A significant multiplicity of
charmed quarks is expected  in heavy-ion collisions at RHIC energies. Their
evolution may carry important signatures if the QGP was a part of the entire
scenario. Here, the theoretical calculations rely heavily on the structure
functions which are taken from  DIS data. The definition of the
heavy flavor DIS structure functions (or sources) should be identical to those
of gluons and light sea quarks.  Only in this way can one consistently
introduce the notion of  intrinsic charm and beauty,  and describe their
excitations  unambiguously. The contribution of heavy flavors to the $ep$-DIS
cross section  should naturally die out when we move in the direction of lower
$Q^2$ along the evolution scale.  However, the status of heavy quarks
in QCD evolution equations has remained uncertain for over two decades. To handle
this problem, one may wish to rely on a minimal extension of the
renormalization group method and account for the heavy quarks only inside the
fermion loop of the gluon self-energy, as is done in Ref.\cite{Doksh}. A kind
of  transient regime  in the vicinity of the  partonic threshold $Q^2\sim
m_c^2$ was suggested in  Refs.~\cite{ACOT,LRSN}. In these calculations, various
regions (below and above the threshold)  were  treated in a different manner: 
with different numbers of active flavors,  and different renormalization 
prescriptions. Special  subtraction schemes were introduced at the threshold. 
However, this method does not eliminate the problem itself: special separate
treatment of the {\em intrinsic}   charm  and beauty, the wee parton inside the
proton, and the {\em extrinsic} charm and beauty,  the heavy quarks created in 
$\gamma^*g$-interaction.  As it will be seen later, both  these
treatments of the massive quark fields in  QCD evolution are incomplete and
it is expedient to consider this problem together with the  problem of the
evolution of static fields.

\section{Deep inelastic electron-proton scattering}
\label{sec:SN2}       

Our approach to high-energy collisions amounts to the elaboration of a  way
to incorporate (to extract and to use) the dynamical information about 
the transient processes which takes place during the collision. Therefore,
we start with the simplest example of  inelastic $ep$-scattering where
at least one of the participants is structureless and pay special
attention to the procedure of measurement. We divide this section into
the three
parts. For the sake of completeness we begin with a brief definition of
the DIS cross-section in terms of quantum field kinetics and define the
null-plane variables which will be used for all the following calculations. 
Next, we discuss the temporal sequence of dynamical processes during
inclusive measurements. Finally, we derive the explicit formula of the
measurement which does not imply the parton decomposition of the proton
and accounts for the static component of the proton's wave function.

\subsection{Observables for DIS}
\label{subsec:SB21}                

Let us consider a collision process with the parameters of only one final-state
particle  explicitly measured. Let this particle be the electron  with 
momentum ${\bf k}'$ and spin $\sigma'$.  The deep inelastic  electron-proton
scattering is an example of such  an experiment.   All the vectors of the
final states which are accepted into the data ensemble are of the form,  
$a^{\dag}_{\sigma'}({\bf k}')|X\rangle $  where the vectors $|X\rangle$  form
a full set. The initial state consists of an electron with  momentum
${\bf k}$ and spin $\sigma$ and some other particle or composite system
carrying quantum numbers $P$.  Thus, the initial state vector is  
$a^{\dag}_{\sigma}({\bf k})|P\rangle $.  The  inclusive transition amplitude
reads as $\langle X|a_{\sigma'}({\bf k}')~S~a^{\dag}_{\sigma}({\bf
k})|P\rangle $   and the inclusive momentum distribution of the final state
electron is the sum of the squared moduli of these amplitudes over the full
set of  non-controlled states $|X\rangle$. Therefore, we obtain the 
following formula, 
\begin{eqnarray} { d N_e\over d {\bf k}'}= \langle
P|a_{\sigma}({\bf k}) S^{\dag} a^{\dag}_{\sigma'}({\bf k}') a_{\sigma'}({\bf
k}') S  a^{\dag}_{\sigma}({\bf k}) |P\rangle ~, \label{eq:E2.1}    
\end{eqnarray}      
which is just an expectation value of the Heisenberg operator of
the number of final state electrons  over the initial state.  Since the
state $|P\rangle$ contains no electrons, one may commute electron creation and
annihilation operators with the $S$-matrix and its conjugate, $~S^{\dag}$,
pulling the Fock operators $a$ and $a^{\dag}$  to the right and to the left,
respectively. Let $\psi_{{\bf k}\sigma}^{(+)}(x)$ be the one-particle  wave
function of the electron. Then the procedure results in the following formula,
\begin{eqnarray} { d N_e\over d {\bf k}'}={1\over 2} \sum_{\sigma\sigma'} 
\int
dxdx'dydy'{\overline\psi}_{{\bf k}\sigma}^{(+)}(x) 
{\overline\psi}_{{\bf k'}\sigma'}^{(+)}(x') \langle P| 
{\delta^2 \over \delta{\overline\Psi}(x)
\delta \Psi(y)} \bigg( {\delta S^{\dag} \over \delta\Psi(y')} 
{\delta S \over \delta {\overline\Psi}(x')}\bigg) |P\rangle  
\psi_{{\bf k}\sigma}^{(+)}(y)\psi_{{\bf k'}\sigma'}^{(+)}(y')~. 
\label{eq:E2.2}    
\end{eqnarray}       
Introducing the Keldysh convention about the contour
ordering \cite{QFK,Keldysh}, we may rewrite this expression as        
\begin{eqnarray} { d N_e \over d {\bf k}'}= { 1 \over 2}
\sum_{\sigma\sigma'}\sum_{AB} (-1)^{A+B} \int dxdx'dydy'{\overline \psi}_
{{\bf k}\sigma}^{(+)}(x) {\overline \psi}_{{\bf k'}\sigma'}^{(+)}(x') 
\langle P|
{\delta^4 S_c \over \delta {\overline\Psi}(x_A) \delta{\overline\Psi}(y'_1)
\delta\Psi(x'_0)\delta\Psi(y_B)} |P\rangle  \psi_{{\bf
k}\sigma}^{(+)}(y)\psi_{{\bf k'}\sigma'}^{(+)}(y')  \label{eq:E2.3}    
\end{eqnarray}     
where $S_c=S^{\dag}S $.  The electron couples only to the
electromagnetic field. Therefore, to the lowest order in electromagnetic
coupling $e$, 
\begin{eqnarray} { d N_e\over d {\bf k}'}={e^2\over 2}
\sum_{\sigma\sigma'} \int dxdy{\overline \psi}_{{\bf k'}\sigma'}^{(+)}(y)
{\overline \psi}_{{\bf k}\sigma}^{(+)}(x) \langle P|\not\!{\bf A}(x)
\not\!{\bf A}(y) |P\rangle  \psi_{{\bf k}\sigma}^{(+)}(y)\psi_{{\bf
k'}\sigma'}^{(+)}(x) ~~, \label{eq:E2.4}     \end{eqnarray}     
where  ${\bf A}(x)$ is the Heisenberg operator of the electromagnetic field.
Already at this very early stage of the calculations, the answer has a clear
physical interpretation. Since only the final state electron is measured, the
probability of the electron scattering is entirely defined by the
electromagnetic field produced by the rest of the system. The final  answer
can be obtained either by using the  equations of QFK  \cite{QGD,QFK}, or one
may take short cut and iterate Eq.~(\ref{eq:E2.4}) by means of the
Yang-Feldman equation,  ~${\bf A}(x)=\int d^4y\Delta_{ret}(x,y){\bf j}(y)$~,
where ${\bf j}(y)$ is the Heisenberg  operator of the electromagnetic current
and $\Delta_{ret}(x,y)$ is the retarded propagator of the photon.    

Summation over the electrons spins brings in the leptonic tensor 
$L_{\mu\nu}(k,k')$.  If $q=k-k'$ is the space-like momentum transfer,
then the DIS cross section is given by
\begin{equation}
 k'_{0} {d\sigma \over d{\bf k'}}= {i\alpha \over(4\pi)^2}
{L_{\mu\nu}(k,k')\over (kP)} 
\Delta_{ret}^{\mu\alpha}(q)W_{\alpha\beta}(q)\Delta_{adv}^{\beta\nu}(q)~~, 
\label{eq:E2.5}
\end{equation}   
where $W^{\mu\nu}(q)$ is the standard Bjorken notation for the 
correlator of  two electromagnetic currents,
\begin{equation}
 W^{\mu\nu}(q)={2V_{lab}P^{0}\over 4\pi}[-i\pi^{\mu\nu}_{10}(q)]~~.
\label{eq:E2.6}\end{equation} 
normalized in a way which provides correspondence to the parton model, and 
%\begin{equation}
$\pi^{\mu\nu}_{10}(x,y)=
-i\langle P|{\bf j}^\mu(x){\bf j}^\nu(y)|P\rangle~$
%\label{eq:E2.7}\end{equation} 
is the electromagnetic polarization tensor, correlator of the currents,
which are the sources of the field which has scattered the electron. 
We accept  the standard  decomposition of the  correlator $W^{\mu\nu}$ in 
momentum space,
\begin{equation}
 W^{\mu\nu}(q)= -e^{\mu\nu}{ W_L \over 2x_{Bj} }+
 \zeta^{\mu\nu}{\nu W_2 \over 2x_{Bj} M^2 }~,
\label{eq:E2.8} \end{equation}
where $\nu = qP$, $Q^2 = -q^2>0$,  $x_{Bj}=Q^2/2\nu$   and
\begin{equation}
e^{\mu\nu}=-g^{\mu\nu}+{q^\mu q^\nu \over q^2 } ; \;\;\;
\zeta^{\mu\nu}=-g^{\mu\nu}+{P^\mu q^\nu+q^\mu P^\nu \over \nu} -
 q^2 {P^\mu P^\nu \over \nu^2 }~.
\label{eq:E2.9}
\end{equation}  
 We shall perform all computations using the infinite momentum
frame fixed by the null-plane vector $n^{\mu}$, ~$n^\mu =(1,{\bf 0_t},-1)$,
~$n^2 =0$.~ It defines the ``+''-components of the Lorentz vectors,
$na = a^+ = 2a_- = a^0 + a^3$;~ $a^- = 2a_+ = a^0 - a^3$.~  
In the infinite momentum frame, the 4-vector of the proton's momentum has
components  ~$P^\mu=(P^+/2,{\bf 0_t},P^+/2)$,~  $P^-=P^0-P^3=0$.
The vector of the momentum transfer has the components,
$~q^\mu=(\nu/P^+,{\bf q_t},-\nu/P^+)$, $~q^+=0$, $~q^-=2\nu/P^+$. 

Instead of the invariant $W_2$, 
we shall use the mass-independent structure function
 $F_2(x_{Bj},Q^2)= \nu W_2 / M^2$,  which is calculated via the equation
\begin{equation}
c_2=  W^{\mu\nu}n_\mu n_\nu ={(P^+)^2 F_2 \over \nu}. 
\label{eq:E2.13}
\end{equation} 

The longitudinal structure function, ~$F_L(x_{Bj},Q^2)=W_L$,~ should be 
calculated in accordance with
\begin{equation}
3F_L = 2x_{Bj}c_1 + 2F_2;\;\;\;\;   c_1 =  W^{\mu\nu}g_{\mu\nu}.    
\label{eq:E2.14}
\end{equation} 

\subsection{Causality and temporal order in DIS}
\label{subsec:SB22}                

The position of the two  photon propagators, $\Delta_{ret}(q)$ and
$\Delta_{adv}(q)$  ( since $\Delta_{adv}$
enters with  inverted space-time arguments,
both are retarded propagators ) in
Eq.~(\ref{eq:E2.5}) reflects the two major aspects  of  causality in
relativistic quantum mechanics.  First, any statement concerning the time
ordering is in manifest agreement with the light cone boundaries.  Second, the
dynamical information about the  quantum-mechanical evolution is read out
only after the evolution is interrupted by the measurement ( the wave
function has  collapsed).  Therefore, the standard inclusive
{\em e-p} DIS delivers
information about quantum fluctuations associated with the proton. The
space-like photon which scatters the electron  ``belongs'' to the
proton and is, in fact, its part. The method of quantum field kinetics
\cite{QFK} allows  one to  continue this line of reasoning  by employing
the basic definition of the electromagnetic polarization tensor
\cite{QFK}. Then, neglecting any corrections to the electromagnetic
vertex, we may rewrite Eq.~(\ref{eq:E2.6}) in the following way,

\begin{equation}
 W^{\mu\nu}(q)=\sum_f e_{f}^{2} {2V_{lab}P_{lab} \over 4\pi}
\int {d^4 p \over (2\pi)^4} {\rm Tr}~ \gamma^\mu {\bf G}_{10}^{f}(p+q)
~\gamma^\nu {\bf G}_{01}^{f}(p)~~.
\label{eq:E2.15} 
\end{equation}
In what follows we shall calculate  only the non-singlet functions of a given
flavor and omit the flavor label $f$  when it causes no confusion. 
The quark field correlators in this equation form a $2\times 2$ matrix and
obey a matrix integral  equation of the Schwinger-Dyson type,
\begin{eqnarray}
 {\bf G}_{AB} = G_{AB}+  G_{AR} \Sigma_{RS} {\bf G}_{SB}~ ~ ~,
\label{eq:E2.15a} \end{eqnarray}         
which allows for a symbolic solution \cite{QFK}. This solution  expresses 
the entire
matrix of  field correlators in terms of the full retarded and advanced
propagators and the sources $\Sigma_{AB}$ (the ``current'' correlators),
\begin{equation}
{\bf G}_{AB} = {\bf G}_{ret}{\stackrel{\leftarrow}{G_{(0)}^{-1}}}
G_{AB} {\stackrel{\rightarrow}{G_{(0)}^{-1}}}{\bf G}_{adv}
+(-1)^{A+B} {\bf G}_{ret} \Sigma_{AB}{\bf G}_{adv}~,  
\label{eq:E2.16} 
\end{equation}
where $ {\stackrel{\rightarrow}{G_{(0)}^{-1}}}$ and
$ {\stackrel{\leftarrow}{G_{(0)}^{-1}}}$  are the differential Dirac 
operators acting  on the right and on the left, respectively. The retarded 
and advanced Green's functions obey more familiar equations,
\begin{eqnarray}
 {\bf G}_{{\stackrel{ret}{\scriptscriptstyle adv}}} = 
G_{{\stackrel{ret}{\scriptscriptstyle adv}}}+  
G_{{\stackrel{ret}{\scriptscriptstyle adv}}} 
\Sigma_{{\stackrel{ret}{\scriptscriptstyle adv}}} 
{\bf G}_{{\stackrel{ret}{\scriptscriptstyle adv}}}~ ~ ~,
\label{eq:E2.17} \end{eqnarray}         
which allow for the symbolic solutions also,
\begin{equation}
 {\bf G}_{{\stackrel{ret}{\scriptscriptstyle adv}}}^{-1} 
= G_{{\stackrel{ret}{\scriptscriptstyle adv}}}^{-1}-
\Sigma_{{\stackrel{ret}{\scriptscriptstyle adv}}}~.
\label{eq:E2.18} \end{equation} 

A simple examination of Eqs.~(\ref{eq:E2.16})
shows that all four dressed correlators ${\bf G}_{AB}$ can be found as
formal solution of the retarded Cauchy problem, with the bare correlators
$G_{AB}$ as the initial data and the self-energies as the sources. Indeed, 
integrating the first term of each these equations twice by parts, we find
for all four elements of ${\bf G}_{AB}$ :
\begin{eqnarray}
{\bf G}_{{\stackrel{10}{\scriptscriptstyle 01}}}(x,y)= 
\int d \Sigma_{\mu}^{(\xi)}
 d \Sigma_{\nu}^{(\eta)} {\bf G}_{ret}(x,\xi)\gamma^{\mu}
G_{{\stackrel{10}{\scriptscriptstyle 01}}}(\xi,\eta)\gamma^{\nu} 
{\bf G}_{adv}(\eta,y) 
-\int d^{4}\xi d^{4} \eta {\bf G}_{ret}(x,\xi)
\Sigma_{{\stackrel{10}{\scriptscriptstyle 01}}}(\xi,\eta) 
{\bf G}_{adv}(\eta,y) ,
\label{eq:E2.18a}
\end{eqnarray} 
\begin{eqnarray}
{\bf G}_{{\stackrel{00}{\scriptscriptstyle 11}}}(x,y)= 
\!\!\int\!\! d\Sigma_{\mu}^{(\xi)}
 d \Sigma_{\nu}^{(\eta)} {\bf G}_{ret}(x,\xi)\gamma^{\mu}
G_{{\stackrel{00}{\scriptscriptstyle 11}}}(\xi,\eta)\gamma^{\nu} 
{\bf G}_{adv}(\eta,y)\! 
-\! \int \!\! d^{4}\xi d^{4} \eta {\bf G}_{ret}(x,\xi)
[\pm G_{0}^{-1}+ \Sigma_{{\stackrel{11}{\scriptscriptstyle 00}}}(\xi,\eta)] 
{\bf G}_{adv}(\eta,y).
\label{eq:E2.18b}
\end{eqnarray}  
               
Though in the original derivation of these equations we  employed the  
``bare'' fields of the interaction picture, the last two formulae allow one
to make a step further and to replace the
correlators $G_{AB}$ of the bare fields with the actual values of
correlators ${\bf G}_{AB}$ on the hypersurface of the initial data which,
in their turn, were created in the course of the preceding evolution.  After
this modification, Eqs.~(\ref{eq:E2.18})--(\ref{eq:E2.18b}) become  an
excellent tool for the study of all inclusive inelastic processes. Similar
equations can be derived for the correlators of the gauge boson fields and
their self-energies.

Viewed in the context of the QFK, the physical picture of the
QCD-evolution looks like the virtual assembling of the source of a proper
electromagnetic fluctuation, one which is capable of providing a
given momentum transfer to the electron. In the event such a
transfer is detected, the hadronic system has to recoil and,
provided the  transferred momentum is high enough, to emit a backward
quark jet. This jet is a successor of the free quark  in the
perturbative vacuum. It is a general principle of the quantum mechanics
of radiation that the emission is the excitation of the field mode. To
be excited, the mode itself should exist. The normal modes of QCD are hadrons
and they have a true physical nonperturbative vacuum as the ground
state. The process of the emission of a free quark is impossible without
creation a new ground state. In the real world of  asymptotically
stable states, the words ``perturbative vacuum'' mean a special highly
excited state of the hadronic system. The QCD evolution includes the
description of its creation by  definition.  Thus, we are unavoidably
led to a physical picture based on the long standing observation by
Shuryak \cite{Shuryak} that the energy scale of confinement is weak,
while the energy density of the QCD vacuum is very high. From this point of
view, multiparticle production in hadronic collisions looks like a
signature of a ``cavitation'' of the physical QCD vacuum and the creation
of a domain with a perturbative  dynamics of quarks and gluons. To
switch it on, one needs extreme Lorentz contraction which resolve
fluctuations of much shorter scale than those providing confinement in
hadrons.  The entire process is strongly localized in space and time
\cite{WD1}. To describe this process  quantitatively, we have to include
two types of correlations:
\begin{eqnarray} 
{\bf G}_{{\stackrel{10}{\scriptscriptstyle 01}}}
=G^{\#}_{{\stackrel{10}{\scriptscriptstyle 01}}} -  
G_{ret}\Sigma_{{\stackrel{10}{\scriptscriptstyle 01}}} G_{adv}~~,~~~ 
%\label{eq:E2.19} \end{eqnarray}     
%\begin{eqnarray} 
{\bf D}_{{\stackrel{10}{\scriptscriptstyle 01}}}
=D^{\#}_{{\stackrel{10}{\scriptscriptstyle 01}}} -  
D_{ret}\Pi_{{\stackrel{10}{\scriptscriptstyle 01}}} D_{adv}~~, 
\label{eq:E2.19a} \end{eqnarray}     
where  the  superscript ``$\#$'' is assigned to all the states in the 
continuum  of the free on-mass-shell fields,
\begin{eqnarray} 
G^{\# ij}_{{\stackrel{10}{\scriptscriptstyle 01}}}(p)  =
-2 \pi i \delta_{ij}(\not\! p +m)  
\theta (\pm p^{0})\delta (p^{2}-m^{2}), \\  \label{eq:E2.20} 
D^{\#ab,\mu \nu}_{{\stackrel{10}{\scriptscriptstyle01}}}(p)  
= -2 \pi i   \delta_{ab} d^{\mu\nu}(p) \theta (\pm p_{0})
\delta(p^{2})~,~~~~ 
d^{\mu\nu}(p)=-g^{\mu\nu}+{p^\mu n^\nu+n^\mu p^\nu \over p^+}~.   
\label{eq:E2.21}\end{eqnarray} 
(Projector $d^{\mu\nu}(p)$ corresponds to the null-plane gauge $A^+=0$.)
These states are initially empty and the vacuum correlators 
$G^{\#}_{10,01}(p)$ and $D^{\#}_{10,01}(p)$ represent only on-mass-shell
quarks and gluons in  the final states.  The second terms in 
Eqs.~(\ref{eq:E2.19a})  include the off-diagonal
self-energies $\Sigma_{{\stackrel{10}{\scriptscriptstyle 01}}}$ 
and $\Pi_{{\stackrel{10}{\scriptscriptstyle 01}}}$, the sources,  which are 
the remainders of the classical configuration of fields in the hadron and 
in the physical
vacuum at some moment of the dynamical evolution before the
final interaction (the measurement).  Before this interaction happens,
the decomposition of the field configuration into the radiation field and
the classical remainders is virtual; the phases of the partial waves are
balanced to form a hadronic correlator propagating through
the condensates of the physical vacuum.

\subsection{Kinematics of the inclusive DIS measurement}
\label{subsec:SB23}  

We do not employ the method of  operator product expansion where $x_{Bj}$
immediately appears as the natural variable corresponding to the light-cone
momentum. Moreover, we are going beyond the parton model and intend to include
new dynamical elements, {\em e.g.}, the static fields, in the evolution equations.
Therefore, we have to pay special attention to the formula of measurement.
In particular, we have to establish a connection between the kinematic
variable $x_{Bj}$ and the Feynman variable $x_F$ without any reference to the
parton model.
The $ep$-DIS measurements rely on a set of rare events where the 
space-like photon $\gamma^*$  is created by a strong change of the state of
one quark only. This quark almost instantaneously decouples from the bulk of
the proton and initiates the jet.  Thus, we have to require that 
one on-mass-shell quark appears in the final state in the perturbative vacuum,
and, therefore, Eq.~(\ref{eq:E2.15}) can be rewritten as
\begin{equation}  
W^{\mu\nu}(q)=\sum_f e_{f}^{2} {2V_{lab}P_{lab} \over 4\pi}
\int {d^4 p \over (2\pi)^4} {\rm Tr} \big[
\gamma^\mu (\not\! p+ \not\! q +m_f)\gamma^\nu
{\bf G}_{ret}(p)\Sigma_{01}(p){\bf G}_{adv}(p) \big]
\big[2 \pi i\delta[(p+q)^2-m^2] \big]~~~.
\label{eq:E2.28}\end{equation}    
By virtue of Eq.~(\ref{eq:E2.13}), 
and introducing $x\equiv x_F=p^+/P^+$, we obtain the expression,
\begin{eqnarray}
F_2(x_{Bj},Q^2) =\sum_f e_{f}^{2}{\nu\over\pi} {V_{lab}P^+ \over 2(2\pi)^3} 
\int_{0}^{1} x d x \int d^2 {\vec p}_t dp^-
\delta[p^+p^- +2\nu x -({\vec p}_t-{\vec q}_t)^2)]~ip^+ {\cal F}^{(f)}(p)~,
\label{eq:E2.29}\end{eqnarray}                                   
where ${\cal F}^{(f)}$ is the coefficient in the 
following decomposition of the quark source,
\begin{eqnarray}
{\bf G}_{ret}\Sigma_{{\stackrel{10}{\scriptscriptstyle 01}}}{\bf G}_{adv}
~=~\not\! p {\cal F}^{{10\choose 01}}(p) +\not\! n p^+ 
{\cal B}^{{10\choose 01}}(p)
+m {\cal C}^{{10\choose 01}}(p)+ 
{\cal H}^{{10\choose 01}}(p)(\not\! p\not\! n-\not\! n\not\! p)~~.
\label{eq:E2.26}\end{eqnarray}   
Following Ref.~\cite{QGD}, we can express the invariant ${\cal F}(p)$ via the
``unintegrated structure function,''
\begin{eqnarray}
{V_{lab}P^+\over 2(2\pi)^3}ip^+{\cal F}^{(f)}(p)=
\delta(p^-){d q_f(x,p_{t}^{2})\over d p_{t}^{2}} ~,
\label{eq:E2.30}\end{eqnarray}
where $\nu= Pq$. [As it will be seen later, in a
certain approximation, $q_f(x,Q^2)$ becomes a structure function of the deep 
inelastic scattering of a- quark of flavor $f$.]  Introduction of the
$\delta$-function  $\delta(p^-)$ at this step of calculations is motivated
physically. According to Eq.~(\ref{eq:E2.5}), the field fragment with the
momentum $p^\mu$ belongs to the yet undestroyed proton.  Let $f(x^+,x^-)\equiv
f(x^0+x^3,x^0-x^3)$ be some proton-related field quantity, {\em e.g.}, the
proper electromagnetic field $A^\mu$ of the proton. If the proton, before its
interaction with the electron, is viewed as a kind of a wave packet that
propagates without dispersion at the speed of light, then the function  $f$
should depend only on a single argument $x^-$. In the momentum representation,
we have
\begin{eqnarray}
f(x^+,x^-)=\int dp^+dp^- f(p^+,p^-) 
e^{-{i\over 2}(p^+x^- +p^-x^+)}~. \nonumber
%\label{eq:E2.31}
\end{eqnarray} 
Thus, in order to eliminate the dependence on $x^+$ and to keep the  proton as
a stable wave packet which propagates along the light cone, one should require 
that  $f(p^+,p^-)= f(p^+)\delta(p^-)$. Admitting the opposite, we would run
into a conflict with the Lorentz contraction of the initial state proton, {\em
viz.}, the radiative corrections to its ``valence structure''  would increase
its size. The dynamics of the quark and gluon fields behind this condition is
totally  nonperturbative.   

In the same way, following Eq.~(\ref{eq:E2.14}), we obtain the expression
for the longitudinal structure function,
\begin{eqnarray}
3F_L(x_{Bj},Q^2) ={ie_{f}^{2}\over\pi} {V_{lab}P^+ \over (2\pi)^3} 
\int_{0}^{1} d x \int d^2 {\vec p}_t dp^- 
\delta[p^+p^- +2\nu x -({\vec p}_t-{\vec q}_t)^2)] {\cal F}_1(p)~,
\label{eq:E2.36}\end{eqnarray}          
where 
\begin{eqnarray}
{\cal F}_1= [2\nu(x_{F}^{2}-x_{Bj}^{2})-x_{Bj}(p^2+m^2)]
{\cal F}-2(p^+)^2 x_{Bj}{\cal B}+4m^2 x_{Bj}{\cal C}.  \nonumber
\end{eqnarray} 
The next calculations are as  follows.  We integrate over the angle
between the vectors ${\vec p}_t$ and  ${\vec q}_t$ and  re-scale
$p_{t}^{2}$ in the remaining integrals by $2\nu$, $y=p_{t}^{2}/2\nu$.
Denoting $\mu=m^2/2\nu$~, we obtain,
\begin{eqnarray}
F_2(x_{Bj},Q^2) ={e_{f}^{2}\over 4\pi} \int_{\mu}^{1} x d x 
\int_{(\sqrt{x_{Bj}}-\sqrt{x-\mu})^2}^{(\sqrt{x_{Bj}}+\sqrt{x-\mu})^2}dy
{d q_f(x,2\nu y)/ d y \over \{[y-(\sqrt{x_{Bj}}-\sqrt{x-\mu})^2]
[(\sqrt{x_{Bj}}+\sqrt{x-\mu})^2-y]\}^{1/2} }~.
\label{eq:E2.32}\end{eqnarray}  
Here, the denominator is the quadrupled area of the triangle with the
sides $p_t$, $q_t$, and $\sqrt{2\nu x - m^2}$ which are rescaled by $2\nu$.
The limits of integration in the Eq.~(\ref{eq:E2.32}) come from the triangle 
inequality.
Now, we should take the limit $\nu\rightarrow\infty$ keeping the $x_{Bj}$
finite. In the new variables, 
the parameter $\nu$ enters only the argument of the
function $d q_f(x, 2\nu y)/dy$.  This function  vanishes in the limit 
of $\nu\rightarrow\infty$. The integral (\ref{eq:E2.32}) vanishes also unless
the variable
$y$ is allowed to take arbitrary small values. Therefore the lower limit of
the integration over  $y$   should be set equal to zero, {\em i.e.}, 
\begin{equation} 
x_F=x_{Bj}+\mu = x_{Bj}(1+m^2/Q^2)~~.
\label{eq:E2.33}
\end{equation} 
For massless quarks, we obtain the usual condition, $x_F=x_{Bj}$. [ If we
were doing the same analysis in the natural momentum variables, then  the lower
limit of the integration over $p_t$ becomes infinite when
$\nu\rightarrow\infty$ at fixed $x_{Bj}$. In fact, the finite interval of
integration over $p_t$ would keep its length $2Q$ and slide as a whole to 
infinitely high values of $p_t$. However, if  Eq.~(\ref{eq:E2.33})  is
satisfied, then both limits  become independent of $\nu$. ] The condition
(\ref{eq:E2.33})  defines the upper limit of the $p_t$-integration integration
also, and the final result reads as
\begin{eqnarray}
F_2(x_{Bj},Q^2) = e_{f}^{2}\bigg( x_{Bj}+{m^2\over 2\nu}\bigg) 
\int_{0}^{4Q^2} d p_{t}^{2}~
 {d~q_f(x_{Bj}+m^2/2\nu,p_{t}^{2})\over d p_{t}^{2} }~~.
\label{eq:E2.35}\end{eqnarray} 
 
Here, the shift of $x_{Bj}$ by  $m^2/2\nu$ is a standard quark mass correction
\cite{Buras}. Change of the upper limit from the familiar $Q^2$ to $4Q^2$ is
unusual but not unexpected. Indeed, of the three sides of the triangle, $p_t$,
$q_t$ and $\sqrt{2\nu x - m^2}$, the third one, by virtue of
Eq.(\ref{eq:E2.33}), is equal to $q_t$. Therefore, the length of the vector
$\vec{p}_t$, the transverse momentum  of the  incoming off-mass-shell field,
varies from $0$ (at ${\vec p}_t={\vec q}_t$)  to $2q_t$ (at  ${\vec p}_t=
-{\vec q}_t$).  Furthermore,  with logarithmic  accuracy there is no 
theoretical difference between $log~Q^2$ and $log~4Q^2$ when
$Q^2\rightarrow\infty$ and no conflict with the leading-$log$ approximation
(LLA), the AP-equations, is anticipated. [For the usual on-mass-shell parton,
we put $\vec{p}_t=0$ and $p^-=0$ and obtain Eq.~(\ref{eq:E2.33}) directly from
the  delta-function in the integrand of Eq.(\ref{eq:E2.29}).]

As it was discussed in Ref.\cite{QGD}, the derivatives of the structure
functions like $dq(x,Q^2)/dQ^2$  (the sources), are similar to the densities 
of states, $~\rho(E)$, of the condensed matter or  scattering theories,  and
the structure functions themselves are similar to the number of states ~$n(E)$~
below the energy $~E$. In an experiment  we measure  $n(E)$ (which is
proportional to the kinematically allowed volume in the phase space),  rather
than  $\rho(E)$. Therefore, it is quite natural that the upper limit of the
$p_t$ integration coincides with the kinematic boundary $2Q$. The less trivial
fact is that the proof of the resonant condition of the measurement,
$~x_{Bj}=x_F$,~  imposes a limit on the rate of the QCD evolution  at high
$Q^2$; the integral over $p_{t}^{2}$ in Eq.~(\ref{eq:E2.29}) should be
convergent. One more important aspect of equations like (\ref{eq:E2.35}) is
that the differential $~d q_f(x,p_{t}^{2})~$ works like a {\em measure} in the
functional space of the of the quantum-mechanical fluctuations of the  quark
field.  Below, a similar kind of measure, $d G(x,p_{t}^{2})$, will be
introduced for the gluon field. However,  the measure will never appear for
those components of the fields which do not have the property of 
propagation and are not independent variables in the Hamiltonian formulation of
the system's dynamics. These are the static components of the fermion field and
longitudinal gluon field.

The full expressions for the invariants ${\cal F}$ and ${\cal F}_1$ in
Eqs.~(\ref{eq:E2.29}) and (\ref{eq:E2.36}) are very long and their exact 
form exceeds the accuracy of approximations used in this paper. The 
complexity comes from the product ${\bf G}_{ret}\Sigma_{01}{\bf G}_{adv}$ in
Eq.~(\ref{eq:E2.28}). Indeed, we have two vectors, $p^\mu$ and $n^\mu$, at our
disposal. The fermion field is massive and none of the fields are polarized.
Thus, all the self-energies $\Sigma_{AB}$ have three terms in their spinor
decomposition,
\begin{equation}
\Sigma(p)=(\not p-m) \sigma_2(p)+ \not n p^+ \sigma_3(p) +m\sigma_m(p)~~.
\label{eq:E2.22}\end{equation}     
 The scalar invariant functions of this decomposition can  be easily
found from the three Dirac traces,
\begin{eqnarray}
{\rm Tr}\Sigma =4m(\sigma_m(p)-\sigma_2(p)),~~~
{\rm Tr}[\not\! n\Sigma]=4p^+\sigma_2(p),~~~
{\rm Tr}[\not\! p\Sigma]=4 [ p^2\sigma_2(p)+(p^+)^2\sigma_3(p)]~.
\label{eq:E2.22a}\end{eqnarray}     
By virtue of Eqs.~(\ref{eq:E2.18}) and (\ref{eq:E2.22}), the retarded
and advanced propagators are readily found as
\begin{eqnarray}
{\bf G}_{{\stackrel{ret}{\scriptscriptstyle adv}}}=
[(\not\! p -m)(1- \sigma_{2}^{{R\choose A}})
-\not\! n p^+ \sigma_{3}^{{R\choose A}}
-m\sigma_{m}^{{R\choose A}}]^{-1}= 
{\cal G}_{{R\choose A}}^{-1}[(1-\sigma_{2}^{{R\choose A}})(\not\! p+m)
-\not\! n p^+ \sigma_{3}^{{R\choose A}}+m\sigma_{m}^{{R\choose A}}]~~,
\label{eq:E2.23}\end{eqnarray}     
where 
\begin{equation}
{\cal G}_{{R\choose A}}=
(p^2-m^2-\sigma_{0}^{{R\choose A}})(1-\sigma_{2}^{{R\choose A}})-
m^2[\sigma_{m}^{{R\choose A}}]^2 
= {\cal S}^{{R\choose A}}_{1}{\cal S}^{{R\choose A}}_{2}
-[{\cal S}^{{R\choose A}}_{m}]^2~~,
\label{eq:E2.24}\end{equation}
and the following short-hand notation is introduced:
$\sigma_0=(p^2-m^2)\sigma_2 +2(p^+)^2\sigma_3+2m^2\sigma_m$,
\begin{eqnarray}
 {\cal S}^{{R\choose A}}_{0}(p)= p^2-m^2 - \sigma_{0}^{{R\choose A}}(p)~,~~  
 {\cal S}^{{R\choose A}}_{2}(p) = 1 - \sigma_{2}^{{R\choose A}}(p)~,~~~
{\cal S}^{{R\choose A}}_{m}(p) =  m\sigma_{m}^{{R\choose A}}(p)~.\nonumber 
%\label{eq:E2.25}
\end{eqnarray} 
For the massless quark field, we can proceed without approximations
and obtain a reasonably simple expression,
\begin{eqnarray}
G_{ret}(p) \Sigma_{01\choose 10}(p) G_{adv}(p)=
{[\not p -\not n(p^2/2p^+)]\sigma_{0}^{01\choose 10}(p)
\over {\cal S}^{R}_{0}(p){\cal S}^{A}_{0}(p)} +{\not n \over 2p^+}
{\sigma_{2}^{01\choose 10}(p)\over 
{\cal S}^{R}_{2}(p){\cal S}^{A}_{2}(p)}~,~~~(~m=0~)
\label{eq:E3.10}\end{eqnarray} 
which explicitly exhibits separation between the dynamical and the constrained
($x^+$--instantaneous or static) parts of the fermion field. The constrained
part (the second term in Eq.~(\ref{eq:E3.10}) ) is easily recognized since
(after the string $G_{ret}\Sigma G_{adv}$ is assembled) it has no pole
corresponding to the propagation. A reasonable approximation for the massive
quark field will be to neglect the radiative corrections in the retarded and
advanced propagators. Then 
\begin{eqnarray}
{\cal F}^{(f)}(p)= {\sigma_{0}^{(f)}\over [p^2-m_{f}^{2}]^{2}}~,~~~
 {\cal B}^{(f)}= -{ \sigma_{3}^{(f)}\over p^2-m_{f}^{2}}~,~~~
{\cal C}^{(f)} ={\sigma_{0}^{(f)}+  \sigma_{m}^{(f)} \over p^2-m_{f}^{2}}~,
\label{eq:E3.11}\end{eqnarray}  
and the expressions for the functions $F_2$ and $F_L$ become significantly
simpler.  The factor $[p^2-m_{f}^{2}]^{-2}$ in the above formulae is never
singular; it originates from the product of the retarded and advanced Green
functions and should be read with the  principal value prescription. 
With this simplified form of the invariants, the integrand of the structure 
function $F_2$, which is  defined by Eq.~(\ref{eq:E2.35}), takes
the following form,
\begin{eqnarray}
\delta(p^-){d q_f(x,p_{t}^{2})\over d p_{t}^{2}}=
{V_{lab}P^+\over 2(2\pi)^3}ip^+{\sigma_{0}^{(f)}(p)\over 
[p^2-m_{f}^{2}]^{2}}~.
\label{eq:E3.15}\end{eqnarray}  
The integrand of the longitudinal structure function 
$F_L$ becomes relatively simple also,
\begin{eqnarray}
{\cal F}_{1}^{(f)}(p)=  
{2m_{f}^{2}~\sigma_{0}^{(f)}(p)\over [p^2-m_{f}^{2}]^{2}} +
{2m_{f}^{2}\over p^2-m_{f}^{2}}\sigma_{m}^{(f)}(p)-\sigma_{2}^{(f)}(p) ~,
\label{eq:E3.13}\end{eqnarray} 
and the final expression can be conveniently written down in the following
form, 
\begin{eqnarray}  
3Q^2F_L(x_{Bj},Q^2) = 
{2m^{2}_{f}Q^2(4Q^2+m^{2}_{f})\over [Q^2+m^{2}_{f}]^2} 
F_2(x_{Bj},Q^2)  \nonumber \\
+{2 e_{f}^{2} x_{Bj}Q^2\over Q^2+m^{2}_{f}}\int_{0}^{4Q^2} d p_{t}^{2}
\bigg[
{d {\cal M}_{f}(x_{Bj}(1+m_{f}^{2}/Q^2),p_{t}^{2})\over d p_{t}^{2} }
+ {\cal Q}_f(x_{Bj}(1+m_{f}^{2}/Q^2),p_{t}^{2}) \bigg]~~,
\label{eq:E3.16}\end{eqnarray}  
where we have introduced the following notation,     
\begin{eqnarray}  
{V_{lab}P^+\over 2(2\pi)^3}ip^+ 
{\sigma_{m}^{(f)}(p)\over  p_{t}^{2}-m_{f}^{2} }=
\delta(p^-){d {\cal M}_f(x,p_{t}^{2})\over d p_{t}^{2}},~~~
{V_{lab}P^+\over 2(2\pi)^3}ip^+ 
\sigma_{2}^{(f)}(p) =
\delta(p^-) {\cal Q}_f(x,p_{t}^{2})~.
\label{eq:E3.17}\end{eqnarray}              
The first term  in Eq.~(\ref{eq:E3.16}) is in direct proportion to $F_2$. 
It is entirely due
to the finite quark mass $m_f$. It shows that at $Q^2\sim m^{2}_{f}$
the non-singlet structure functions $F_2$ and $F_L$ of the heavy quark
may be of the same order.

\section{Evolution of the sources}
\label{sec:SN3}       

Our next step is to find evolution equations for the sources of the
electromagnetic fluctuation which scatters the electron. These sources 
are given by the off-diagonal self-energies $\Sigma_{01}$ (for the quarks)
and $\Sigma_{10}$ (for the anti-quarks). Their general expressions were
derived in Refs.~\cite{QFK,QGD} and, as in Ref.~\cite{QGD}, we shall use
approximation of the bare vertices. Furthermore, we shall use
simplified equations (\ref{eq:E2.19a}) which define 
evolution of the quark sources and make similar
simplifications for the gluons (hereafter, the gauge $A^+=0$ for the gluon
field is assumed).
% \begin{eqnarray}
%{\bf D}_{{\stackrel{01}{\scriptscriptstyle 10}}}=
%D^{\#}_{{\stackrel{01}{\scriptscriptstyle 10}}}+
%{\bf D}_{ret}\Pi_{{\stackrel{01}{\scriptscriptstyle 10}}} 
%{\bf D}_{adv}~, 
%\label{eq:E3.1}
%\end{eqnarray}    

The general expression for the gluon sources, $\Pi_{01\choose 10}$, were
derived in Refs.~\cite{QFK,QGD} also and an approximation of the bare vertices
will be used here as well,
\begin{eqnarray}
 \Sigma_{01 \choose 10}(p)=ig_{r}^{2} C_F \!\!
\int \!\! {d^4 k \over (2\pi)^4}
 \{ \gamma_\mu {\bf G}_{ret}(k) 
\Sigma_{01\choose 10}(k) {\bf G}_{adv}(k)
 \gamma_\nu D^{\#\nu\mu}_{10\choose 01}(k-p) \nonumber \\
+ \gamma_\mu G^{\#}_{01 \choose 10}(k+p) \gamma_\nu
\left[{\bf D}_{ret}(k) \Pi_{10 \choose 01}(k) 
{\bf D}_{adv}(k)\right]^{\nu\mu}\} 
\label{eq:E3.3}\end{eqnarray}   
\begin{eqnarray}
  \Pi^{\mu\nu}_{01}(p)=-ig_{r}^{2} 
\{-\int \! {d^4 k \over (2\pi)^4}  {\rm Tr}\{
\gamma^{\mu}{\bf G}_{ret}(k) \Sigma_{01}(k) {\bf G}_{adv}(k)
 \gamma^{\nu} G^{\#}_{10}(k-p) 
+\gamma^{\mu} G^{\#}_{01}(k+p)\gamma^{\nu}
{\bf G}_{ret}(k) \Sigma_{01}(k) {\bf G}_{adv}(k)] \nonumber \\
+{1\over 2}\int  {d^4 k \over (2\pi)^4}
  V^{\mu\alpha\nu}_{acf}(p,k-p,k)
\left[{\bf D}_{ret}(k) \Pi_{01}(k) {\bf D}_{adv}(k)\right]^{\alpha\beta}_{cc'}
 V^{\nu\beta\sigma}_{bc'f'}(-p,p-k,-k) 
 D_{10, f'f}^{\#\lambda\sigma}(k-p) \}~. 
\label{eq:E3.4}\end{eqnarray} 
           
The renormalization of these equations as well as the study of their infra-red
behavior  is a special problem.  The idea of renormalization in
the QFK-based calculations has been discussed in Ref.~\cite{QGD} for the case
of pure glue-dynamics.  In this simplest case the  coupling constant should be
replaced by the ``running coupling,''   $g_{r}^{2}\rightarrow 4\pi
\alpha_s(p_{t}^{2})$.  However, we still have not obtained satisfactory balance
of the collinear singularities except for the limiting 
case when the evolution is
described by the DGLAP equations. The status of this problem will be discussed
separately \cite{HQ2}.

\subsection{Splitting of the evolution equations}
\label{subsec:SB31}  

The polarization tensor $\Pi^{\mu\nu}$  only appears between the retarded and
advanced propagators, {\em i.e.}, in the combination
$[D_{ret}(p)\Pi (p)D_{adv}(p)]^{\mu\nu}$.
Both propagators contain the same projector, $d^{\mu\nu}(p)$
which (by the gauge condition) is orthogonal to the 4-vector $n^{\mu}$. So, of
the general tensor form, only two terms survive,
\begin{equation}
 \Pi^{\mu\nu}(p) =g^{\mu\nu}~p^2~w_1(p)+ p^\mu p^\nu w_2(p).
\label{eq:E3.5}\end{equation}  
The others, like $p^{\mu}n^{\nu}+ n^{\mu}p^{\nu}$  or $n^{\mu}n^{\nu}$ ,
will cancel out.  Introducing one more projector,
\begin{equation}
{\overline d}^{\mu\nu}(p)=- d^{\mu\rho}(p) d_{\rho}^{\nu}(p) 
=-g^{\mu\nu}+{p^\mu n^\nu+n^\mu p^\nu \over (np) }
 -p^2{n^\mu n^\nu \over  (p^+)^2} ,
\label{eq:E3.6}\end{equation}          
which is orthogonal to both vectors $n^{\nu}$ and $p^{\mu}$ , we find that 
the invariants $w_1$ and $w_2$ can be found from two convolutions,
\begin{eqnarray}
-{\overline d}_{\mu\nu}(p) \Pi^{\mu\nu}(p) =2p^2~ w_1(p),~~~and~~~
n_\mu n_\nu \Pi^{\mu\nu}(p)=(p^+)^2~ w_2(p)~,
\label{eq:E3.6a}\end{eqnarray} 
independently of the other invariants accompanying the missing tensor 
structures.  The new projector, which includes only two transversal
gluon modes, naturally appears in the tensor with a gluon source.
Indeed, solution of the Schwinger-Dyson equation for the retarded gluon 
propagator,
\begin{equation}
{\bf D}_{{\stackrel{ret}{\scriptscriptstyle adv}}} 
= D_{{\stackrel{ret}{\scriptscriptstyle adv}}}
+  D_{{\stackrel{ret}{\scriptscriptstyle adv}}} 
\Pi_{{\stackrel{ret}{\scriptscriptstyle adv}}} 
{\bf D}_{{\stackrel{ret}{\scriptscriptstyle adv}}}~.
\label{eq:E3.2} \end{equation} 
 can be cast in the form,
\begin{eqnarray}
{\bf D}^{\mu\nu}_{ret\choose adv}(p)=
{{\overline d}^{\mu\nu}(p)\over p^2 - p^2 w_{1}^{R\choose A}(p)}+
{1 \over (p^+)^2}~ {n^{\mu}n^{\nu} \over  1 - w_{2}^{R\choose A}(p)}. 
\label{eq:E3.7}\end{eqnarray}   
With the shorthand notation,
%\begin{eqnarray}
 $~{\cal W}^{{R\choose A}}_{1}(p) = p^2 - p^2~w_{1}^{R\choose A}(p)$, and
 $~{\cal W}^{R\choose A}_{2}(p) = 1 - w_{2}^{R\choose A}(p)$, 
%\label{eq:E3.8}\end{eqnarray}     
we easily obtain,
\begin{eqnarray}
\left[D_{ret}(k) \Pi_{01\choose 10}(k) D_{adv}(k)\right]^{\mu\nu}=
{-{\overline d}^{\mu\nu}(p)p^2 w_{1}^{01\choose 10}(p)\over 
{\cal W}^{R}_{1}(p){\cal W}^{A}_{1}(p)}+
{ w_{2}^{01\choose 10}(p)n^{\mu}n^{\nu}
\over (p^+)^2{\cal W}^{R}_{2}(p){\cal W}^{A}_{2}(p)}.
\label{eq:E3.9}\end{eqnarray}   
Once again, similar to the case of the fermion field, we encounter
the longitudinal ($x^+$-instantaneous) part $w_2$ of the gluon
source  which is not accompanied by the propagation poles in the string 
$D_{ret}\Pi D_{adv}$.  

The role of the tree propagators  ${\cal S}^{R\choose A}$ and   ${\cal
W}^{R\choose A}$ which include the radiative corrections  is clearly
understood.  These corrections keep memory about the balance of the phases
between  various fields in the yet undestroyed proton. In fact, they {\em
nonperturbatively} maintain the $\delta(p^-)$-prescription, the requirement
that the proton does not fall apart in the course of the QCD evolution. 
Keeping this important qualitative observation in mind we shall limit
ourselves, in what follows, to the bare retarded and advanced (tree)
propagators of the quark and gluon fields.  

Our next goal is to find the evolution equations for
the three invariants, $\sigma_{0}^{(f)}(x,p_{t}^{2})$,
$\sigma_{L}^{(f)}(x,p_{t}^{2})$ and  $\sigma_{m}^{(f)}(x,p_{t}^{2})$.
Using the Eqs.~(\ref{eq:E2.22a}) and (\ref{eq:E3.6a}) we easily extract
evolution equations for various invariants from the tensor evolution
equations (\ref{eq:E3.3}) and (\ref{eq:E3.4}).  In sequence, we obtain
three equations for three invariants of the fermion source,
\begin{eqnarray}
\sigma_{0}^{(f)}(p)={2 g^2(p_{t}^{2})\over (2\pi)^3}
\int d^4 k \bigg\{ C_F \delta_+[(k-p)^2]   \nonumber \\   
\times \bigg[\bigg( \big[-{p^2-m_{f}^{2}\over z }+(k^2-m_{f}^{2})\big] 
{z^2+1\over 1-z} - 2 m_{f}^{2} \bigg)
{\sigma_{0}^{(f)}(k) \over [k^2-m_{f}^{2}]^2 } 
+z \sigma_{2}^{(f)}(k) - 2m_{f}^{2}(1-z)
{\sigma_{m}^{(f)}(k)\over k^2-m_{f}^{2}} \bigg]  \nonumber \\ 
+ \delta_+[(k-p)^2-m_{f}^{2}] \bigg[
\big[\big(-{p^2-m_{f}^{2}\over z }+k^2\big){(1-z)^2+z^2\over 2}
+m_{f}^{2}\big]k^2{w_{1}^{(10)}(-k)\over [k^2]^2} - 
z(1-z) w_{2}^{(10)}(-k)\bigg]\bigg\}~~,
\label{eq:E3.18}\end{eqnarray}    
\begin{eqnarray}
\sigma_{2}^{(f)}(p)={2 g^2(p_{t}^{2})\over (2\pi)^3}
\int d^4 k \bigg\{C_F \delta_+[(k-p)^2]
~{1\over z}~ {\sigma_{0}^{(f)}(k)\over [k^2-m_{f}^{2}]^2}  
 + \delta_+[(k-p)^2-m_{f}^{2}] ~{1-z\over 2z}~
{k^2 w_{1}^{(10)}(-k)\over [k^2]^2} \bigg\},
\label{eq:E3.19}\end{eqnarray}     
\begin{eqnarray}
\sigma_{m}^{(f)}(p)={2 g^2(p_{t}^{2})\over (2\pi)^3}
\int d^4 k \bigg\{ C_F \delta_+[(k-p)^2]
\bigg[{1-z\over z}~ {\sigma_{0}^{(f)}(k)\over [k^2-m_{f}^{2}]^2} 
 -  {\sigma_{m}^{(f)}(k)\over k^2-m_{f}^{2}}  \bigg] \nonumber \\ 
+\delta_+[(k-p)^2-m_{f}^{2}]~ {1\over 2z}~
{k^2w_{1}^{(10)}(-k)\over [k^2]^2} \bigg\}.  
\label{eq:E3.20}\end{eqnarray} 
and two equations for the invariants of the gluon source,
\begin{eqnarray}
p^2~w_{1}^{(01)}(p)=-{2 g^2(p_{t}^{2})\over (2\pi)^3}
\int d^4 k \bigg\{ T_f \sum_{f}
\delta_+[(k-p)^2-m_{f}^{2}]  \nonumber  \\
\times \bigg[  \big[(-{p^2 \over z}+ k^2-m_{f}^{2}) 
{(z-1)^2+1\over z} -2 m_{f}^{2} \big]
{\sigma_{0}^{(f)}(k)\over [k^2-m_{f}^{2}]^2}+(1-z)\sigma_{2}^{(f)}(k)
-2m_{f}^{2} z {\sigma_{m}^{(f)}(k)\over k^2-m_{f}^{2}} \bigg]  \nonumber\\
- N_c \delta_+[(k-p)^2] \bigg[ \big(-{p^2 \over z}+k^2\big)
\bigg({1-z\over z}+{z\over 1-z}+z(1-z)
\bigg)k^2{w_{1}^{(01)}(k)\over [k^2]^2} - 
~(z-{1\over 2})^2 w_{2}^{(01)}(k)\bigg]\bigg\}~~,
\label{eq:E3.21}\end{eqnarray}                                     
\begin{eqnarray}
w_{2}^{(01)}(p)={2 g^2(p_{t}^{2})\over (2\pi)^3}
\int d^4 k \bigg\{4 T_f \sum_{f}
\delta_+[(k-p)^2-m_{f}^{2}]
~{1-z\over z^2}~{\sigma_{0}^{(f)}(k)\over [k^2-m_{f}^{2}]^2} \nonumber\\  
-N_c \delta_+[(k-p)^2]  ~(1 -{1\over 2z})^2~
{k^2 w_{1}^{(01)}(k)\over [k^2]^2}\bigg\} ~.
\label{eq:E3.22}\end{eqnarray}     
Here, the sum over $(f)$ runs over all quark flavors  and anti-flavors. The
longitudinal fields appear in the evolution equations  
(\ref{eq:E2.16})--(\ref{eq:E2.19a}) in an alternating regime. This
remarkable feature of the  evolution equations  has a very clear physical
explanation. Indeed, if the static source has emitted a propagating wave,
at least one additional emission is necessary to create a new static
field configuration.

Deriving the field equations (\ref{eq:E2.16})--(\ref{eq:E2.19a}) we were
relying upon the most straightforward implementation of the idea of quantum
mechanical evolution; {\em viz.}, the Heisenberg picture of the evolution of 
observables of  DIS with the quark and gluon fields of a stable hadron  in
physical vacuum as the initial data.  Eqs.~(\ref{eq:E3.18})--(\ref{eq:E3.22})
are the mathematical expression of this physical picture. The initial data are
not given explicitly, and if they  were, the range of the incorporated physical
information should be the same as in the local operators of the OPE. However,
we do not view the perturbative part of 
the evolution as the scattering problem and
this results in the   new feature of the above equations; the invariants
$\sigma_2$ and $w_2$ which correspond to the sources of longitudinal fields
participate in the evolution equations on the same footing as the sources 
$\sigma_0$ and $w_1$ of the transverse fields.  Therefore, we have a reason to
suspect that the entire dynamics of the QCD evolution carries a  classical
component. This is in compliance with the qualitative understanding of the QCD
evolution as the virtual decomposition of the hadrons, which are the genuine
fundamental modes of  QCD, in terms of the alien set of modes defined as
excitations above an artificial perturbative vacuum.  The parton picture should
emerge after certain approximations, different from the formal twist
classification of the local operators of the OPE expansion. 

\subsection{Feed-back via longitudinal fields}
\label{subsec:SB32}  

In what follows, we shall make several approximations and  try to estimate
their accuracy. In the OPE-based calculations, it is a formidable  task.
Indeed, the hierarchy of scales in OPE is two-fold. On the one hand, the
operator functions are ordered by their twist which, would we wish to
treat it statistically, corresponds to the order of the irreducible
correlation  function in condensed matter physics. On the other hand, the
coefficient functions are ranked by the order of their expansion in powers
of the coupling constant. Because of the capricious  interplay of  these two
essentially different expansions it is difficult to obtain a reliable 
estimate of the accuracy even for the widely used DGLAP equations.  

We shall pose the  problem in a different way and try to 
understand what elements of the  physical
picture are lost when the field equations (\ref{eq:E3.18})--(\ref{eq:E3.22})
are approximated by the master equations of the parton evolution.   
Let us consider the transverse components of the sources, $\sigma_0$ and 
$w_1$, as granted and use the connection to the 
observables, equations (\ref{eq:E3.13})--(\ref{eq:E3.17}), 
\begin{eqnarray}
c ~ip^+{\sigma_{0}^{(f)}(p)\over 
[p^2-m_{f}^{2}]^{2}}=  \delta(p^-)
{d q_f(x,p_{t}^{2})\over d p_{t}^{2}}~,~~~
c~ip^+\sigma_{2}^{(f)}(p)=  \delta(p^-) {\cal Q}_f(x,p_{t}^{2})~,
\label{eq:E3.23}\end{eqnarray}
for the fermion field, and introduce the similar links for the gluon field,   
\begin{eqnarray}
c~ip^+{p^2w_{1}(p)\over [p^2]^{2}}= 
 \delta(p^-){d G (x,p_{t}^{2})\over d p_{t}^{2}}~,~~~
c~ip^+~w_{2}(p)= 
\delta(p^-){\cal G} (x,p_{t}^{2})~,
\label{eq:E3.24}\end{eqnarray}   
where $c$ is an insignificant common normalization constant. Integrating
Eqs.~(\ref{eq:E3.19}) and (\ref{eq:E3.22}) over $p^-$ and using the
delta-functions to integrate the variable $k^-$ out, we obtain,
\begin{eqnarray}
{\cal Q}_f(x,p_{t}^{2}) =
c~\int dp^-~ip^+\sigma_{2}^{(f)}(p)={g^2(p_{t}^{2})\over (2\pi)^3}
\bigg[C_F \int_{p^+}^{P^+} {d k^+ \over k^+-p^+} q_f (k^+) +
{1\over 2} \int_{p^+}^{P^+} {d k^+ \over k^+} G (k^+)\bigg] ~,
\label{eq:E3.25}\end{eqnarray}       
\begin{eqnarray}
{\cal G}(x,p_{t}^{2}) =
c~\int dp^-~ip^+w_{2}(p)={g^2(p_{t}^{2})\over (2\pi)^3}
\bigg[ 4 T_f\sum_{f} \int_{p^+}^{P^+} {d k^+ \over p^+} q_f (k^+) -
N_c \int_{p^+}^{P^+} {d k^+ \over k^+-p^+}~{(2p^+-k^+)^2\over 4 p^+k^+}
G (k^+)\bigg] ~,
\label{eq:E3.26}\end{eqnarray}       
where 
\begin{eqnarray}
q_f(p^+)=\int d{\vec p}_t {d q_f(x,p_{t}^{2})\over d p_{t}^{2}}~~~~  
{\rm and}~~~~ 
G(p^+)=\int d{\vec p}_t {d G(x,p_{t}^{2})\over d p_{t}^{2}}~,
\label{eq:E3.27}\end{eqnarray}        
are the $x$-fractions of the quarks and the glue converted into the
radiation, integrated over all transverse momenta. The requirements for the
convergence of these integrals are exactly the same as in the proof of the
resonant condition for the measurement, Eq.~(\ref{eq:E2.33}).  The physical
meaning of these equations can be uncovered by examining , {\em e.g.}, 
the firsty 
term in Eq.~(\ref{eq:E3.26}), the rate of the depleting of the initial
reservoir of static glue is proportional to the total number of  quarks
radiated above a given value $x_F$. The  collinear divergence at  $k^+=p^+$
in  Eqs.~(\ref{eq:E3.25}) and (\ref{eq:E3.26}) is not due to the 
spurious pole of the gluon propagator. It appears as a consequence of 
the singularity of the
integration $dp^-$ of the isolated mass shell delta function,
$\delta[(k^+-p^+)(k^--p^-)-({\vec k}_t-{\vec p}_t)^2]$,  at the point
$k^+=p^+$.  This singularity  appears only for  the longitudinal modes. It
is shielded when the delta-function is accompanied by the  propagators of
the transverse field with the off-mass-shell space-like momentum, {\em
i.e.}, when the field is slowed down by its virtuality.  In coordinate space,
this singularity corresponds to the integration of the $x^+$-independent
function in the infinite limits, and it is unavoidable when the wave packet
is moving without dispersion in the light-like direction, thus depending
only on $x^-$. To shield this singularity, one has either to decelerate
the proton (to take $P^+\sim \sqrt{s}$~ finite), or to establish a lower
limit of resolution for the transverse momentum. In the second case, the
natural scale of hadronic confinement, $\Lambda_{QCD}$, has to be taken as
the cut-off.  Both cut-offs play the same role; they allow for the
geometrical separation of the fields between the near and far zones.  
Physically, in QCD one cannot separate two waves if their momenta differ
by less than the width of the first peak of  diffraction on an object with
the size  $\sim\Lambda_{QCD}^{-1}$. In any case, integration near the lower
limit will invoke a large logarithm, $ log~(s/\Lambda_{QCD}^{2})$, which
will over-compensate for the smallness of the coupling constant.  However, we
must realize that this singularity is a direct consequence of the
unphysical geometry of the fields in the infinite momentum frame; the cut-off
for $P^+$, though quite understandable physically,  is alien to the
mathematical formulation which employs a fraction $x_F$ of the light-cone
momentum as the main variable.

More accurate estimate of the cut-off in the splitting kernels can be
obtained if the process is viewed in the scope of the wedge form of
dynamics \cite{WD1,WDG}. This dynamics employs the gauge condition
$A^\tau=0$ and it has the  gauge $A^+=0$ as its limit in vicinity of the
null-plane $x^-=0$, when $\tau\to 0$ and $\eta\to -\infty$.  Here, $\tau$
and $\eta$ parameterize the space-time coordinates in $tz$-plane,~
$t=\tau\cosh\eta,~~z=\tau\sinh\eta$.

For our immediate purpose, it is enough
to consider only one component of the on-mass-shell gluon correlator,
\begin{eqnarray} 
D_{10}^{00}(x_1,x_2)=\int {d{\vec k}_t d\theta \over 2(2\pi)^3}~
{4\sinh\eta_1\sinh\eta_2 {\vec k}_{t}^{2}e^{-ik(x_1-x_2)}
\over (k^+ e^{-\eta_1}+k^- e^{\eta_1})(k^+ e^{-\eta_2}+k^- e^{\eta_2}) }
\approx  \int {d{\vec k}_t d\theta\over 2(2\pi)^3}~
{ k^- e^{-ik(x_1-x_2)}
\over k^+ +( e^{2\eta_1}+ e^{2\eta_2})(k_{t}^{2}/k^+) }
\label{eq:E3.27a}\end{eqnarray} 
where $k^+=k_te^\theta$, $k^-=k_t e^{-\theta}$ and $k^+k^-=k_{t}^{2}$.
Now, it is easy to see that the pole $1/k^{+}$ is shielded. The cut-off
is defined by the location of the interaction region; for the lowest 
$k^+$ one should take $\eta_{1,2}\sim -Y/2
\approx -\ln(2E_{max}/m_{hadr}) \approx\ln(\sqrt{s}/\Lambda_{QCD})$. 
The invariant energy of the collision, $\sqrt{s}$, or its equivalent,
 the full width $Y$ of the hadronic rapidity plateau, is the measure of
the resolution in longitudinal direction of the interaction which initiate
the deeply inelastic process. Any extension of the theory beyond the
parton model will include them explicitly.     

The next step is to examine the feed-back which appears when $\sigma_2$
and $w_2$ from Eqs.~(\ref{eq:E3.25}) and (\ref{eq:E3.26}) are inserted
into the right hand side of the evolution equations (\ref{eq:E3.18}) and 
(\ref{eq:E3.21}) for the propagating (transverse) components of the quark
and gluon fields. Since these terms in the integrand
depend on $k_t$ only via the coupling constant, all integrations except 
the one over $k^+$ can be carried out, {\em viz.},
\begin{eqnarray} 
I_{q\to q}=
\int dp^- dk^- d{\vec k} {\delta_+[(k-p)^2]\delta(k^-) 
g_{r}^{2}(k_{t}^{2})\over [p^2-m^2]^2}
=\int {g_{r}^{2}(({\vec k}+{\vec p})^{2}) d{\vec k} 
\over [zk_{t}^{2}+(1-z)(p_{t}^{2}+m^2)]^2}
\approx {\pi g_{r}^{2}(p_{t}^{2}) k^+ \over p^+ (p_t^2+m^2)}~,\nonumber\\ 
I_{q\to g}
%=\int dp^-dk^-d{\vec k}{\delta_+[(k-p)^2-m^2]\delta(k^-)\over [p^2-m^2]^2}
={\pi g_{r}^{2}(p_{t}^{2}) k^+ \over p^+}{1-z \over m^2+(1-z) p_t^2 }~,~~~ 
I_{g\to g}
%=\int dp^-dk^-d{\vec k}{\delta_+[(k-p)^2]\delta(k^-)\over [p^2]^2}
={\pi g_{r}^{2}(p_{t}^{2}) k^+ \over p^+ ~p_t^2}~,~~~ 
I_{g\to q}
%=\int dp^-dk^-d{\vec k}{\delta_+[(k-p)^2-m^2]\delta(k^-)\over [p^2]^2}
={\pi g_{r}^{2}(p_{t}^{2}) k^+ \over p^+}{1-z \over z m^2+(1-z) p_{t}^{2}}~, 
\label{eq:E3.28}\end{eqnarray}   
for the ``feed-back'' for the propagating fields via the longitudinal
modes from quarks to quarks, quarks to glue, glue to glue and  glue to
quarks  , respectively. At $m=0$, all these terms yield a simple and
remarkable function,
\begin{eqnarray}
{g_{r}^{2}(p_{t}^{2})  \over p^+ p_t^2} \int k^+ ~f(k^+)~dk^+~, \nonumber
\end{eqnarray} 
which contains a universal Weizsacker-Williams denominator  corresponding
to the static field of the ultra-relativistic charge. This {\em induced}
static source is given, in its turn, by the integral over the light-cone
momenta of the transverse fields above the currently probed momentum
$p^+$.  Almost mysteriously, this dependence appears  for the  fermion  
field also; it corresponds to the static pattern of the fermion wave 
function!

The intensities of the induced static sources are almost independent of the
transverse  momenta. It may be tempting to treat these sources as the
equivalents of the valence quarks and gluons. However, there is a major
difference. Evolution of the valence partons has to begin through splitting
kernels, and this is the key point in derivation of the {\em master} AP
equation;  while in the field approach advocated here, the longitudinal modes
are coupled  to  the transverse ones in a special way which regenerate the
distribution of the field inherent to  the ultra-relativistic classical source!
Moreover, in this approach, the evolution ladder is built from the above and
only from the above, solely  from the requirement of the given momentum
transfer in the measurement and in agreement with the  causal picture of the
inclusive measurement. This points us to the most important observation that
the classical field pattern is inherent to the light front evolution,
regardless what were the initial data.

The feed-back fragment of the field evolution equations can be cast in the
following form,
\begin{eqnarray}
\bigg\{{d q_f(x,p_{t}^{2})\over d p_{t}^{2}}\bigg\}_{f.b.}=
{\pi g^2(p_{t}^{2})\over (2\pi)^3}
\int_{p^+}^{P^+} {d k^+ \over k^+}~\bigg[  C_F
 {z\over p_t^2+m_{f}^{2}}~{\cal Q}_f (k^+) -  
{z(1-z)^2\over m_{f}^{2}+(1-z) p_t^2 }{\cal G} (k^+)\bigg]~,
\label{eq:E3.29}\end{eqnarray}   
\begin{eqnarray}
\bigg\{{d G (x,p_{t}^{2})\over d p_{t}^{2}}\bigg\}_{f.b.}=
{\pi g^2(p_{t}^{2}) \over (2\pi)^3}
\int_{p^+}^{P^+} {d k^+ \over k^+}~\bigg[
-N_c {(z-1/2)^2 \over p_t^2}~{\cal G} (k^+) -  T_f \sum_f
{(1-z)^2\over z m_{f}^{2}+(1-z)p_t^2}{\cal Q}_f (k^+)\bigg]~.
\label{eq:E3.30}\end{eqnarray}   
where the subscript {\em f.b.}  stands for the additional contribution to 
the rate of evolution of the transverse fields from the feed-back via the
longitudinal ones. ${\cal Q}$  and ${\cal G}$ stand for  the left side of
Eqs. (\ref{eq:E3.25}) and (\ref{eq:E3.26}) which connect longitudinal
sources to the intensity of the radiated transverse fields. This
contribution can be estimated in the following way. [From here to
the end of this section we shall neglect the quark masses.]

The data, parameterized via DGLAP equations, clearly indicate that the 
quantity $xq(x,Q^2)$ for the sea quarks experiences a certain growth at
small $x$. Let us start with the most conservative  probe function,
~$q_f(x)=\kappa/x$,~ and calculate the feed-back along the string
$q_f(x)\to {\cal G}(x) \to G(x)$. Using Eqs. (\ref{eq:E3.25}) and 
(\ref{eq:E3.29}) we obtain the first estimate,
\begin{eqnarray}
\bigg\{ {p_{t}^{2}\over g^2(p_{t}^{2}) }~
{d G(x,p_{t}^{2})\over d p_{t}^{2}} \bigg\}_{f.b.}^{q_f\to {\cal G} \to G} 
= -{4 \pi T_f N_c g^2 \over (2\pi)^6} \int_{x}^{1} {d y \over y}~ 
\bigg[{1\over 4}-{x\over y}+{x^2\over y^2}\bigg] 
\int_{y}^{1} {d u \over y}~{\kappa\over u} \nonumber \\
=- {4 \pi T_f N_c g^2 \kappa \over (2\pi)^6} 
\bigg[{1\over 12} {1\over x}~ \ln {1\over x}-{1\over 9}~{1\over x}
+ {\cal O}(1)\bigg]~~, 
\label{eq:E3.30a}\end{eqnarray}           
which clearly indicates that our probe function is bad.  The feed-back
response,~ $\sim(1/x)\ln(1/x)$,~ is  much bigger than the probe signal,
~$\sim 1/x$ ~! This is exactly what happens in the calculations of 
the first  order
quantum correction to the Weizsacker-Williams field \cite{Raju95}. Let us
take another probe function, one which exhibits more realistic
behavior, {\em e.g.},  $q_f(x)=\kappa x^{-1-\lambda}$. Repeating the same
calculations, we obtain,
\begin{eqnarray}
\bigg\{ {p_{t}^{2}\over g^2(p_{t}^{2}) }~
{d G(x,p_{t}^{2})\over d p_{t}^{2}} \bigg\}_{f.b.}^{q \to {\cal G} \to G} 
=- { \pi T_f N_c g^2 \over (2\pi)^6} 
{\lambda^2+\lambda +2 \over \lambda(\lambda +1)(\lambda +2)(\lambda +3)} 
 ~~{\kappa\over x^{1+\lambda}} + {\cal O}(x^{-1})~~, 
\label{eq:E3.30b}\end{eqnarray}    
Thus, our second probe is very good. The low-$x$ dependence  $\sim 
x^{-1-\lambda}$ appears to be an eigen-function of the feed-back via the
longitudinal fields!  To confirm this statement, let us examine the
feed-back along the string $G(x)\to {\cal G}(x) \to G(x)$, taking 
$G(x)=\gamma~x^{-1-\lambda}$ as the probe function: 
\begin{eqnarray}
\bigg\{ {p_{t}^{2}\over g^2(p_{t}^{2}) }~
{d G(x,p_{t}^{2})\over d p_{t}^{2}} \bigg\}_{f.b.}^{G \to {\cal G} \to G} 
=+ {\pi N_{c}^{2} g^2 \gamma \over (2\pi)^6} \int_{x}^{1} {d y \over y}~ 
\bigg[{1\over 4}-{x\over y}+{x^2\over y^2}\bigg] 
\int_{y}^{1} {d u \over u^{1+\lambda}} \bigg({1\over 4y}-{1\over u}+
{1\over 4(u-y)} \bigg)~~,
\label{eq:E3.30c}\end{eqnarray}           
Retaining the terms which are most singular at low $x$, we obtain the
estimate,
\begin{eqnarray}
\bigg\{ {p_{t}^{2}\over g^2(p_{t}^{2}) }~
{d G(x,p_{t}^{2})\over d p_{t}^{2}} \bigg\}_{f.b.}^{G \to {\cal G} \to G} 
=+ {\pi N_{c}^{2} g^2  \over (2\pi)^6} 
\bigg[ {1-3\lambda\over \lambda(\lambda +1)}+ 
 \ln{s\over \Lambda^2} \bigg]
{\lambda^2+\lambda +2 \over 16(\lambda +1)(\lambda +2)(\lambda +3)} 
~~{\gamma \over x^{1+\lambda}}~~,
\label{eq:E3.30d}\end{eqnarray}     
which supports the self-similarity of the power-like enhancement at low $x$
with respect to the feed-back. This dependence holds both for quarks and
gluons. 
If $~\ln(s/\Lambda^2)~$ is considered to be parametrically large, then the
leading terms are as follows, 
\begin{eqnarray}
\bigg\{ {p_{t}^{2}\over g^2(p_{t}^{2}) }~
{d q(x,p_{t}^{2})\over d p_{t}^{2}} \bigg\}_{f.b.} 
= {\pi  g^2(p_{t}^{2})  \over (2\pi)^6} 
{1 \over x^{1+\lambda}}~\ln{s\over\Lambda^2}~
\bigg[ {C_{F}^{2} \over \lambda +2}~\kappa 
+{ N_c \over 4(\lambda +2)(\lambda +3)}~\gamma\bigg]~~,
\label{eq:E3.30e}\end{eqnarray}           
\begin{eqnarray}
\bigg\{ {p_{t}^{2}\over g^2(p_{t}^{2}) }~
{d G(x,p_{t}^{2})\over d p_{t}^{2}} \bigg\}_{f.b.} 
= {\pi g^2(p_{t}^{2})  \over (2\pi)^6} 
{1 \over x^{1+\lambda}}~\ln{s\over\Lambda^2}~
\bigg[-{C_F T_f n_f \over (\lambda +1)(\lambda +2)}~\kappa 
+{N_{c}^{2}(\lambda^2+\lambda +2 )
\over 4(\lambda +1)(\lambda +2)(\lambda +3)} ~\gamma  \bigg]~.
\label{eq:E3.30f}\end{eqnarray}               
         
The feed-back is positive along the string ~$G \to {\cal G} \to G$~ and
negative along the strings involving quarks, like ~$q \to {\cal G} \to G$,
~$q \to {\cal Q} \to G$,~ {\em etc.}. The gluons tend to boost the rate of
their own evolution, while the quarks slow  the gluon evolution down.
Evolution of quarks is always boosted by the feed-back.  Though
Eqs.~(\ref{eq:E3.30d})-(\ref{eq:E3.30f}) contain an infra-red regulator,
$~\ln{s/ \Lambda^2}$,~ which is inconsistent both with the philosophy and
the technical design of the infinite momentum frame, it looks as though the
potential source of the low-$x$ enhancement in deep inelastic processes is
found correctly.  The QCD evolution of observables in high-energy
inclusive processes includes the evolution of classical (longitudinal) quark
and  gluon fields which can be responsible for the enhancement.  The 
power-like behavior at low $x$ has been predicted 
long ago by Balitsky, Fadin, Kuraev, and Lipatov from the solution of the
BFKL equation \cite{BFKL} and the exponent $\lambda$ was explicitly 
found in the
case of pure glue-dynamics.  Considerable work is needed to find this 
exponent in our approach. Furthermore, it is not yet clear if  these
two approaches rely on exactly the same physical input.

Our conclusion about the power-like enhancement at low $x$ is very close to 
that of McLerran {\em et al.}~\cite{Larry} but it is motivated in a
different way. First, we neither employ nor even need the valence quarks as
the  classical source
of the gluon field. In fact, we keep in mind that the strongest gluon fields
are due to the vacuum condensates which exist even in the absence of
hadrons. In the state of confinement, neither hadrons nor nuclei drag the
glue; they propagate through it. Second, the coupling between the static 
and the propagating fields only weakly depends  on transverse momentum,
only
via $\alpha_s(p_{t}^{2})$.  Stronger dependence might come from the
propagators, but the longitudinal modes do not propagate. The dynamics of the
measurement always leads to  {\em induced} static fields which have
a steep behavior at low $x$ but are integrated over all $p_t$. For this
reason, accurate measurement of the longitudinal structure function $F_L$,
which is  directly connected to the static component of the quark field,
is of extreme importance.  
                                                                
The  classical pattern of the light-front evolution poses a severe problem.
It is well known that  energy-momentum conservation can be obtained
 either in quantum or in classical theory of radiation.
Quantum theory in the presence of classical external field does not allow for
the consistent formulation of the momentum conservation
since even the transverse classical field can
be presented at most as a superposition of states with various numbers of
quanta in every mode (coherent states). Longitudinal fields are always
``external'' and they are not even a subject for quantization. Therefore, we
have to admit that the equations of the light front QCD evolution cannot
have the first integral of the light-cone momentum.  However, 
momentum conservation plays an important role in the derivation of the
AP evolution equations, being solely responsible for the so-called
$+$-prescription which regulates the collinear singularities of the
splitting kernels. With momentum conservation the parton model
is well motivated. Otherwise, we face the problem of identifying the
subject of the QCD evolution itself.

\subsection{Evolution of classical fields and the structure function $F_L$}
\label{subsec:SB33}    

To leading order,
the standard  evolution equations completely disregard the dynamics of
the longitudinal fields in the QCD evolution,
thus missing a physically important part of the dynamical process.  
As it follows from the expression (\ref{eq:E3.16})
for the longitudinal structure function $F_L$, the function ${\cal Q}$
is solely responsible for the scaling violation of $F_L$ when the quarks are
massless.
Eq.~(\ref{eq:E3.25}) indicates that ${\cal Q}$ depends on $p_t$ only
via the running coupling $~g_r(p_{t}^{2})\sim 1/log(p_{t}^{2}/\Lambda^2)$.
Therefore, it is easy to estimate the $Q^2$-dependence of the non-singlet
longitudinal structure function $F_L$ for the case
of massless quarks, 
\begin{eqnarray}
Q^2F_L\sim\int_{0}^{4Q^2}{d p_{t}^{2} \over \ln(p_{t}^{2}/\Lambda^2)}=
\Lambda^2 ~{\rm li}({4Q^2\over\Lambda^2}) 
= -\Lambda^2~E_1\big(-\ln({4Q^2\over\Lambda^2})\big)
\approx {4Q^2 \over \ln(4Q^2/\Lambda^2)} \sim Q^2\alpha_s(Q^2)~,
\label{eq:E3.30g}\end{eqnarray}
where ${\rm li}(x)$ is the integral logarithm.  
Therefore, $F_L$ is not strongly suppressed  ({\em
i.e.}, by the powers of $\Lambda^2 / Q^2$)
with respect to $F_2$.  In the standard OPE approach, the
non-singlet $F_L$ appears only in the next-to-leading order. Indeed, if
the longitudinal fields were eliminated from the initial data by putting
the partons on mass shell in a perturbative vacuum, at least one extra
emission is needed to regenerate the longitudinal component at  the level
of the  coefficient functions. For this reason, the OPE-based result
contains additional $Q^2$-dependence due to the scaling violation of the
$F_2(x,Q^2)$ and  $G(x,Q^2)$. The OPE-based predictions for large $x$,
when one might expect the strongest effect from the longitudinal fields,
systematically lie below the data. In our case,  according to
Eqs.~(\ref{eq:E3.25})--(\ref{eq:E3.27}), the geometry of the static fields 
depends on the overall  radiation above the currently probed value of the
$x_F$; the entire range of $Q^2$ is integrated. 
An active involvement of the longitudinal (classical) fields in the 
QCD evolution explains the enhancement of the longitudinal structure function
$F_L$ due to the quark masses, the first two terms in 
Eq.~(\ref{eq:E3.16}); the heavier the quark, the more static is its
field.

\section{Master equations of QCD evolution}
\label{sec:SN4}  

Equations of motion of the quark and gluon fields do not allow one to
eliminate longitudinal fields from the evolution equations by fiat. 
At most,  one can use  a renormalization procedure which absorbs
their effects into the definition of some composite objects. The simplest
objects of this kind are  known as the particles in asymptotic states. 
Once this normalization is done, the only consistent way to proceed is to
treat the entire problem as a problem of scattering. The initial state
should be introduced as an asymptotic one. Thus, we are compelled to
introduce the parton picture. The proton has  to be presented as a  system
of free (valence?) quarks long before its interaction with the electron.
Even the concept of ``before'', which is so fruitful in the inclusive
approach, becomes ambiguous.  Instead of an initial picture of the inclusive
{\em e-p} interaction, we must use the picture of the electron-parton
scattering. In fact, we restrict ourselves to the study of the interaction
between the electron and a single parton  which is defined as an
elementary excitation above the perturbative vacuum.  The initial state of
this  parton has to be prepared. This is done implicitly by introducing a 
factorization scale. In fact, the factorization scale $\mu_{0}^{2}$ undermines
the basic idea of  the inclusive measurement in {\em e-p} scattering. To
comply with quantum mechanics, this parameter has to be explicitly
measured. Moreover, this measured value  has to be used in theoretical
calculations!  The recently discovered HERA events with  rapidity gaps
are good candidates to allow for this kind of measurements.

This qualitative analysis prompts further approximations. In order
to obtain the master equations, $\sigma_2$ and $w_2$ must be  eliminated
from the equations (\ref{eq:E3.18})--(\ref{eq:E3.22}) without any 
discussion of the accuracy. 
After that, one can
express $\sigma_0$, $\sigma_m$, and $w_1$ via the observables of DIS,
and integrate the equations over $p^-$. We end up with two integral 
equations for quarks,
\begin{eqnarray}
\bigg\{{d q_f(x,p_{t}^{2})\over d p_{t}^{2}} \bigg\}_{LL}
={g^2(p_{t}^{2})\over (2\pi)^3}  \int_{x}^{1} {dy \over y}
\int d^2 {\vec k} \bigg\{ C_F  \bigg[\bigg( 
{1\over H({\vec p},{\vec k},z)+(1-z) m_{f}^{2}}
~{z^2+1\over 1-z}  \nonumber \\           
- z {(k_{t}^{2}+m_{f}^{2})(1+z^2)+
2m_{f}^{2}(1-z)\over [H({\vec p},{\vec k},z)+(1-z) m_{f}^{2}]^2}
\bigg)~{d q_f(y,k_{t}^{2})\over d k_{t}^{2}}  
-{ 2m_{f}^{2} z (1-z)^2 \over [H({\vec p},{\vec k},z)+(1-z) m_{f}^{2}]^2}~
{d {\cal M}_f(y,k_{t}^{2})\over d k_{t}^{2}}\bigg]  \nonumber \\    
+ \bigg[ {(1-z)^2+z^2\over 2}
\bigg( {1\over H({\vec p},{\vec k},z)+ m_{f}^{2}} -
{z(1-z)~k_{t}^{2}\over [H({\vec p},{\vec k},z)+ m_{f}^{2}]^2}\bigg)
+{ m_{f}^{2}~z~(1-z)
  \over [H({\vec p},{\vec k},z)+ m_{f}^{2}]^2}\bigg]
{d G (y,k_{t}^{2})\over d k_{t}^{2}}  \bigg\}~~,  
\label{eq:E3.31}\end{eqnarray}    
\begin{eqnarray}
\bigg\{{d {\cal M}_f(x,p_{t}^{2})\over d p_{t}^{2}} \bigg\}_{LL}
={ g^2(p_{t}^{2})\over (2\pi)^3}  
\int_{x}^{1} {dy \over y} \int d^2 {\vec k} 
\bigg\{ {C_F \over  H({\vec p},{\vec k},z)+(1-z) m_{f}^{2}}
\bigg[-(1-z)~{d q_f(y,k_{t}^{2})\over d k_{t}^{2}} 
+ z ~{d {\cal M}_f(y,k_{t}^{2})\over d k_{t}^{2}}\bigg] \nonumber \\    
- {1\over 2 (H({\vec p},{\vec k},z)+ m_{f}^{2})}
{d G (y,k_{t}^{2})\over d k_{t}^{2}}  \bigg\}~~.  
\label{eq:E3.32}\end{eqnarray} 
and one equation for gluons,
\begin{eqnarray}
\bigg\{{d G(x,p_{t}^{2})\over d p_{t}^{2}} \bigg\}_{LL}
= { g^2(p_{t}^{2})  \over  (2\pi)^3}\int_{x}^{1} {dy \over y}
 \int d^2 {\vec k} 
\bigg\{ T_f \sum_{f} \bigg[
\bigg( \bigg({-1\over H({\vec p},{\vec k},z)+ zm_{f}^{2}}
+{z(1-z)(k_{t}^{2}+m_{f}^{2})
\over [H({\vec p},{\vec k},z)+ z  m_{f}^{2}]^2 }\bigg)
{(1-z)^2+1\over z}   \nonumber\\ 
+{ 2z(1-z)  m_{f}^{2}
\over [H({\vec p},{\vec k},z)+ z  m_{f}^{2}]^2 }  \bigg)
~{d q_f(y,k_{t}^{2})\over d k_{t}^{2}} 
+{ 2z^2(1-z)  m_{f}^{2}
\over [H({\vec p},{\vec k},z)+ z  m_{f}^{2}]^2 }~ 
{d {\cal M}_f(y,k_{t}^{2})\over d k_{t}^{2}}\bigg]  \nonumber\\ 
+ N_c \bigg( {1-z\over z}+{z\over 1-z}+z(1-z)\bigg)
\bigg[{1\over H({\vec p},{\vec k},z)}-
{(1-z)k_{t}^{2} \over H^2({\vec p},{\vec k},z)}\bigg]
{d G (y,k_{t}^{2})\over d k_{t}^{2}}  \bigg\}~~.
\label{eq:E3.33}\end{eqnarray}  
where ~$H({\vec p},{\vec k},z)=z({\vec p}-{\vec k})^2 +(1-z)p_{t}^{2}~$.  
The subscript LL stands for the Leading Logarithms, the destination point
of our study.  It indicates that the rate of the evolution is
considered with the  switched off static fields. The further simplification 
of these equations will result in the DGLAP equations.
One can easily see that Eq.~(\ref{eq:E3.32}) describes evolution of
the dynamical 
quark mass (in fact, ${\cal M}$ is the imaginary part of the ``pole mass''
which is defined by the retarded self-energy of the quark). 
Negative signs on the right side of Eq.~(\ref{eq:E3.32}) 
indicate that evolution of the effective mass leads to  a faster decrease, 
the higher the $Q^2$  which is probed. This trend clearly supports 
our intuitive understanding
of the interplay between the quark mass and the transverse momentum transfer. 
By examination, Eq.~(\ref{eq:E3.32}) is of the same type as the other
two evolution equations and the solution for $~{\cal M}_f~$
must be the standard logarithmic
exponent. There is no reason to neglect $~{\cal M}_f(x,p_{t}^{2})$, unless
$~m_{f}^{2}\ll p_{t}^{2}$.

Equations (\ref{eq:E3.31})--(\ref{eq:E3.33}) are still more complicated
than the DGLAP equations. Unlike the  DGLAP equations, they are the  {\em
integral equations  for the rates of evolution} and exhibit non-locality
in transverse directions. The kernels of these equations are singular;
except for the familiar spurious poles, $z=1$,  of the splitting
functions, the nonlocal part of the kernels contains overlapping infrared
singularities at $z=1$ and ${\vec k}={\vec p}$. As discussed above, this
behavior is quite consistent with the crude picture of the infinite momentum
frame. Its smoothing requires  a physical cut-off, which can be natural
only in a more gentle picture.  The cut-off  brings into play new large
logarithms, like ~$log(s/\Lambda^2)$,~ which are  alien to the dynamics of
the  infinite momentum frame and cannot really be estimated by its
internal means.  Even $x_F$, the main variable of the parton model, has to be
sacrificed in order to incorporate the energy of the collision as a
physical parameter of the theory. In fact, this parameter is vitally
needed in order to bring into the theory the Lorentz contraction of the 
colliding hadrons as the measure of the initial resolution in the longitudinal
direction.

To eliminate the need for the infrared cut-off, once again, by ordinance,
one should follow the strategy of hunting the leading logarithms, either
$log(Q^2/\Lambda^2)$, or $log(1/x)$.  As it was shown in the previous  
section, it is inconsistent to address the low-$x$ behavior without
explicit account of the longitudinal fields (however, they are already 
neglected in Eqs.~(\ref{eq:E3.31})--(\ref{eq:E3.33}) ).
In order to keep track of the leading logarithms  $~log(Q^2/\Lambda^2)$,
one should find a
solution to the scattering problem in the Born's approximation and
exponentiate it. This is known also as ordering by angles or ordering of the
emission by the transverse momenta. To obtain the same result from
Eqs.~(\ref{eq:E3.31})--(\ref{eq:E3.33}), it is necessary to eliminate
non-local effects in their kernels (by neglecting $k_t$ with respect to
$p_t$) and to introduce the upper limit $p_{t}^{2}$ for the  integration
$dk_{t}^{2}$. The integration itself becomes straightforward and yields,
\begin{eqnarray}
\bigg\{{d q_f(x,p_{t}^{2})\over d p_{t}^{2}} \bigg\}_{LL}
={g^2(p_{t}^{2})\over (2\pi)^3}  \int_{x}^{1} {dy \over y}\bigg\{ 
C_F  \bigg[\bigg( {1\over p_{t}^{2}+(1-z) m_{f}^{2}}~{z^2+1\over 1-z}  
-{m_{f}^{2}~z[2+(1-z)^2] \over [p_{t}^{2}+(1-z) m_{f}^{2}]^2}
\bigg)~q_f(y,p_{t}^{2})   \nonumber \\  
-{ 2m_{f}^{2} ~z ~(1-z)^2 \over [p_{t}^{2}+(1-z) m_{f}^{2}]^2}~
{\cal M}_f(y,p_{t}^{2})\bigg] + \bigg( {(1-z)^2+z^2\over 2} 
{1\over p_{t}^{2}+ m_{f}^{2}}+{m_{f}^{2}~z~(1-z) 
  \over [p_{t}^{2}+ m_{f}^{2}]^2}\bigg) G (y,p_{t}^{2})  \bigg\}~~,  
\label{eq:E3.34}\end{eqnarray}    
\begin{eqnarray}
\bigg\{{d {\cal M}_f(x,p_{t}^{2})\over d p_{t}^{2}} \bigg\}_{LL}
={ g^2(p_{t}^{2})\over (2\pi)^3}  \int_{x}^{1} {dy \over y} 
\bigg\{ {C_F \over  p_{t}^{2}+(1-z) m_{f}^{2}}
\bigg[-(1-z)~ q_f(y,p_{t}^{2}) + z {\cal M}_f(y,p_{t}^{2})\bigg]  
- {G(y,p_{t}^{2}) \over 2 (p_{t}^{2}+ m_{f}^{2})}  \bigg\}~~.  
\label{eq:E3.35}\end{eqnarray} 
\begin{eqnarray}
\bigg\{{d G(x,p_{t}^{2})\over d p_{t}^{2}} \bigg\}_{LL}
= { g^2(p_{t}^{2})  \over  (2\pi)^3}\int_{x}^{1} {dy \over y}
\bigg\{ T_f \sum_{f} \bigg[\bigg( \bigg({-1\over p_{t}^{2}+ zm_{f}^{2}}
+{z(1-z)m_{f}^{2}\over [p_{t}^{2}+ z  m_{f}^{2}]^2 }\bigg)
{(1-z)^2+1\over z}   \nonumber\\ 
+{ 2z(1-z)  m_{f}^{2}\over [p_{t}^{2}+ z  m_{f}^{2}]^2 }  \bigg)
~q_f(y,p_{t}^{2}) +{ 2z^2(1-z)  m_{f}^{2}
\over [p_{t}^{2}+ z  m_{f}^{2}]^2 }~  {\cal M}_f(y,p_{t}^{2})\bigg]
+ N_c \bigg( {1-z\over z}+{z\over 1-z}+z(1-z)\bigg)
{G (y,p_{t}^{2})\over p_{t}^{2}}~ \bigg\}~~.
\label{eq:E3.36}\end{eqnarray}  

Thus, the angular ordering results in drastic simplification of the
evolution equations. Only the splitting kernels $P_{qq}$ and $P_{gg}$ are
singular now.  However, we cannot regularize these singularities by means
of the (+)-prescription as is done in AP equations. Indeed, in the AP approach
the (+)-prescription maintains two conservation laws, conservation of the
light cone charge, $~j^+$, and conservation of the  light-cone component
of the momentum, $~p^+$, in the process of  sequential splitting of the
parton.  The parton is supposed to be free and no interaction terms are
included into the operators $~j^+$ and $~p^+$. Though the dynamical  mass
${\cal M}_f$ of the fermion does not enter these operators, they are
independent partners of the parton densities $q$ and $G$ in the evolution
equations.  Unless $m_f=0$,  the  conservation laws do not follow from the
(+)-prescription.  This fact has a very clear physical interpretation.
Though the wave equation for the massive quark is of hyperbolic type and
allows for the propagation of signals along the light-cone characteristics,
these signals cannot correspond to the on-mass-shell states of heavy
quarks.  Only dynamical polarization waves which include  massive 
fermion fields can propagate at speed of light. Even though we can include the
effect of quark masses into the gluon propagator  and, eventually, into
the running coupling, this will lead to an  excess of  accuracy if 
the dynamical mass term ${\cal M}_f$  is not included into the evolution 
equations.

Our result for the running coupling \cite{HQ2}
coincide with the one obtained by
Dokshitser and Shirkov \cite{Doksh} ,
\begin{eqnarray}
 {1\over g_r^2(\mu^2)}-{1\over g_r^2(M^2)} ={1\over 16\pi^2}
 ~\bigg[-N_c[{11\over 6}
%-2\ln~{\mu^+\over \epsilon}
]~\ln{M^2\over \mu^2} 
 +\sum_fT_f{1\over3}\big[\ln{M^2\over \mu^2}+
f({\mu^2\over m_f^2})-f({M^2\over m_f^2})\big]\bigg]~.
\label{eq:B50}
\end{eqnarray}  
where the function $f$ is obtained from the one-loop gluon self-energy with
the subtraction at an arbitrary space-like momentum $p^2<0$,
\begin{eqnarray}
f({p^2\over m^2}) = \ln{-p^2\over 2m^2}-{4m^2\over -p^2}+
\big(1-{2m^2\over -p^2}\big)
  \sqrt{1+{4m^2\over -p^2}} 
~\ln{\sqrt{1+{4m^2\over -p^2}}-1
\over \sqrt{1+{4m^2\over -p^2}}+1}~,
\label{eq:B38}\end{eqnarray} 
and has the following limiting behavior in  different
domains of the transverse momentum,
\begin{eqnarray}
f({p^2\over m^2})\sim {\cal O}({m^2\over p^2})~,~~~|{p^2\over m^2}~|\gg ~1;~~~~~~
f({p^2\over m^2})\to \ln{-p^2\over 2m^2}~,~~~ |{p^2\over m^2}|~\ll~ 1. 
\label{eq:B37}
\end{eqnarray}
These formulae smoothly interpolate running coupling between the domains with
different transverse momenta and provide different numbers of ``active
flavors''.  
However, it is important to emphasize that with the quark masses retained, 
the dynamical quark mass should be kept in the evolution equations and
we still face problem of the
infrared regularization of the evolution equations.

\section{Summary}

The method of QFK allows one to extend the standard definition of the evolution
equations beyond the scope of the parton model. The extended equations include
new elements, the static (longitudinal) gluon field and the static
(non-propagating) quark fields. The quark masses are accounted for also and the
dynamical mass term in the evolution equations is sensitive to the change of
scale in the same way as the quark and gluon structure functions. Though
the role of the new terms is perfectly understood physically, the extended set
of equations suffers from the infrared-singular terms which cannot be regulated
in the the standard scheme where the proton is analyzed in the infinite
momentum frame.

\vskip 1cm
\centerline {\bf ACKNOWLEDGMENTS}

\bigskip 

We are grateful to Ramona Vogt who critically read the manuscript and to L.
McLerran, A. Vainshtein, R.Venugopalan and H. Weigert for many stimulating
discussions.  We thank the [Department of Energy's] Institute for Nuclear
Theory at the University of Washington for its hospitality and support during
the completion of this work.

This work was supported by the U.S. Department of Energy under Contract 
No. DE--FG02--94ER40831.

\end{document}